\pdfoutput=1
\documentclass[11pt]{article}
\usepackage[hmargin=2.54cm,vmargin=2.54cm]{geometry} % see geometry.pdf
\usepackage{geometry}
\usepackage{amsmath,amssymb,setspace,fancyhdr,geometry,url,color}
\usepackage{subfigure,graphicx,caption}
\usepackage{lscape,amsthm}
\usepackage[authoryear,round]{natbib}
\usepackage{tabularx}
\usepackage{arydshln}
\usepackage{multirow}
\usepackage{hhline}
\RequirePackage{lineno} 
\allowdisplaybreaks[0]
\allowbreak
\title{Changes in Spatio-temporal Precipitation Patterns in Changing Climate Conditions}
\author{Won Chang, Michael L. Stein, Jiali Wang, V. Rao Kotamarthi, Elisabeth J. Moyer}

\begin{document}

\maketitle
\doublespacing
%%%%%%%%%%%%%%%%%%%%%%%%%%%%%%%%%%%%%%%%%%%%%%%%%%%%%%%%%%%%%%%%%%%%%
% ABSTRACT
%
% Enter your Abstract here

\abstract{\small Climate models robustly imply that some significant change in precipitation patterns will occur. Models consistently project that the intensity of individual precipitation events increases by approximately 6-7\%/K, following the increase in atmospheric water content, but that total precipitation increases by a lesser amount (1-2\%/K in the global average in transient runs). Some other aspect of precipitation events must then change to compensate for this difference. We develop here a new methodology for identifying individual rainstorms and studying their physical characteristics -- including starting location, intensity, spatial extent, duration, and trajectory -- that allows identifying that compensating mechanism. We apply this technique to precipitation over the contiguous U.S.\ from both radar-based data products and high-resolution model runs simulating 80 years of business-as-usual warming. In model studies, we find that the dominant compensating mechanism is a reduction of storm size. In summer, rainstorms become more intense but smaller; in winter, rainstorm shrinkage still dominates, but storms also become less numerous and shorter duration. These results imply that flood impacts from climate change will be less severe than would be expected from changes in precipitation intensity alone. We show also that projected changes are smaller than model-observation biases, implying that the best means of incorporating them into impact assessments is via ``data-driven simulations'' that apply model-projected changes to observational data. We therefore develop a simulation algorithm that statistically describes model changes in precipitation characteristics and adjusts data accordingly, and show that, especially for summertime precipitation, it outperforms simulation approaches that do not include spatial information.
}

\section{Introduction}

Some of the most impactful effects of human-induced climate change may be changes in precipitation patterns \citep[AR5;][]{ipcc2013ipcc}. 
Changes in future flood and drought events and water supply 
may incur social costs and mandate changes in management practices 
\citep[e.g.][]{rosenzweig1994potential,vorosmarty2000global,christensen2004effects,barnett2008human,karl2009global,nelson2009climate,piao2010impacts}. Understanding and projecting changes in future precipitation patterns is therefore important for informed assessment of climate change impacts and design of adaptation strategies.

\begin{figure}
\begin{center}
\includegraphics[scale=0.6]{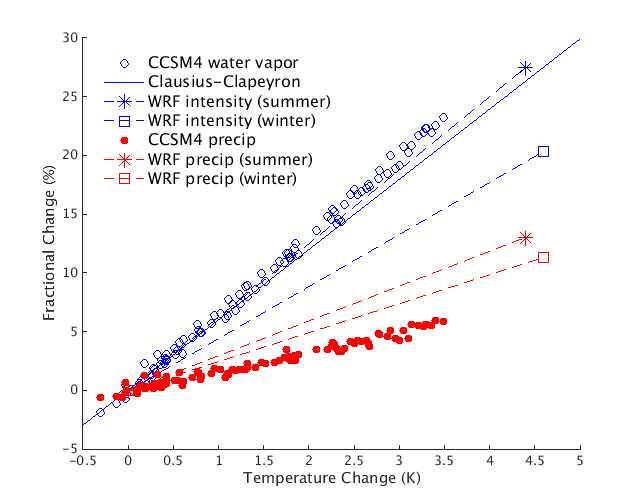}
\end{center}
\caption{Comparison of projected changes in precipitation amount (red) and in intensity or water vapor content (blue). Circles show annual values of global mean output from a CCSM4 run from the CMIP5 archive under the RCP 8.5 scenario; here open blue circles are water vapor. Blue line shows theoretical saturated water content at global mean temperature given by the Clausius-Clapeyron relation. Dashed lines show output over the continental U.S. from the CCSM4/WRF regional climate runs described in Section 2 and used throughout this work. Here blue lines show precipitation intensities. Changes in summer and winter seasons are shown separately (asterisks and squares, respectively). In all cases, total precipitation rises much less steeply than do precipitation intensity or water vapor content.}
\label{figure:puzzle}
\end{figure}

%Original Text: Future changes in spatio-temporal precipitation patterns are not well studied, but climate models show consistent hydrological changes in simulations under higher CO$_2$ \citep[e.g.][]{hennessy1997changes,giorgi2005updated,tebaldi2004regional}, which imply that some changes must occur.
{\color{black} Future changes in spatio-temporal precipitation patterns are not well studied, but climate simulations under higher CO$_2$ consistently show divergent increases in precipitation amount and intensity that, if true, imply that some changes must occur.} %reviewer1 specific1
Models robustly project uniform increases in precipitation intensity (amount per unit time and unit area when rain occurs) of $\approx 6\%$ per degree temperature rise, following atmospheric water content governed by Clausius-Clapeyron \citep[e.g.][]{held2006robust,willett2007attribution,stephens2008controls,wang2012review}. However, model total precipitation rates rise differently: in transient runs, in the global average, by only 1-2\% per degree \citep[e.g.,][]{knutson1995time,allen2002constraints,held2006robust,stephens2008controls,wang2012review}. 
 (Time-averaged precipitation rise can exceed Clausius-Clapeyron only in the deep tropics.) 
Nearly all latitudes then show a discrepancy between changes in rainfall intensity and total amount that must be ``compensated'' by some other change in precipitation characteristics \citep[e.g.][]{hennessy1997changes}.
In the midlatitudes, the amount/intensity discrepancy generally resembles the global average 
(Figure \ref{figure:puzzle}). %, which shows model projections over the contiguous U.S.\ from the Community Climate System Model 4 (CCSM4) and from regionally downscaled runs used in the study). XX save this for caption
%XX leave this for caption runs from the CMIP5 archive \cite{WCRP2015CMIP5}. Similar to the findings in other studies, the increase in precipitation in this model run (about 2\%/K) is much lower than the increase in intensity when it does rain (about 7\%/K). 
Model midlatitude rain events must therefore experience changes in frequency (fewer storms), duration (shorter storms), or size (smaller storms).

Current approaches to studying future precipitation cannot determine changes in rainstorm characteristics, because they do not consider individual events. Studies generally analyze precipitation at individual model grid cells \citep{knutson1995time,semenov2002secular,stephens2008controls} or over broad regions \citep{tebaldi2004regional,giorgi2005updated}. Effects from changes in rainstorm frequency, duration, and size are all confounded in local or spatially-aggregated time series. To overcome this limitation, we need an approach that identifies, tracks, and analyzes individual rainstorms.
Currently-used rainstorm tracking algorithms are not appropriate for climate studies: they are designed for severe convective storms in the context of nowcasting and forecasting using weather forecasting models \citep{davis2006object2}, radar \citep[e.g.][]{dixon1993titan,johnson1998storm,wilson1998nowcasting,fox2005bayesian,xu2005kernel,davis2006object1,han20093d,lakshmanan2009efficient}, or satellite images \citep{morel2002climatology1,morel2002climatology2}. Storm studies in climate projections have also focused only on very large and intense events \citep{hodges1994general,cressie2012dynamical}. These algorithms cannot efficiently handle lower-intensity precipitation events with more complicated morphological features and evolution patterns. Studying future precipitation patterns requires new storm identification and tracking strategies.
%Therefore, we need a new algorithm that can also find and track rainstorms with more complex behavior.

Such studies also require high-resolution model output. For a model to plausibly represent changes in real-world rain events it must at minimum explicitly represent those events.  Computational constraints mean that typical General Circulation Model (GCM) runs used for climate projections have spatial resolution of order 100 km, too coarse to represent morphological features of localized rainstorms. 
Fine-grained observational data can be generated from radar datasets;  
studies of future precipitation require dedicated model runs at similar resolution, at the price of more limited time series. 

Finally, making practical use of insights into changes in rainstorm characteristics requires methods for combining model projections with observational data. 
%is to generate future precipitation by combining model output and observational data. 
Hydrological and agricultural impact assessments cannot use scenarios of future precipitation from even high-resolution models, since model precipitation  can differ considerably from that in observations \citep[e.g.][]{ines2006bias,baigorria2007assessing,teutschbein2012bias,muerth2013need}. % XX are all these citations for coarse GCMs? 
%Impacts assessments using raw model output can be highly misleading. 
The two primary current approaches to addressing these biases are bias correcting model output based on observations (of means or marginal distributions) \citep[e.g.][]{ines2006bias,christensen2008need,piani2010statistical1,piani2010statistical2,teutschbein2012bias} and ``delta'' methods that adjust observations by model-projected changes (in means or marginal distributions) \citep[e.g.][]{hay2000comparison,raisanen2013projections,raty2014evaluation}. Neither approach allows representing changes in spatio-temporal dependence. %For bias correction approaches, 
Some work has extended bias correction methods to also address biases in spatio-temporal dependence \citep{vrac2015multivariate}, but these methods do not allow for future \textit{changes} in that dependence. %Existing delta methods capture current present-day spatio-temporal dependence but do not address future changes. 
An important objective of this work is therefore to extend delta methods to capture model-projected changes in rainstorm characteristics, while still ensuring the greatest fidelity to real-world precipitation statistics.

In this work we both seek to understand the changes in rainstorm characteristics in future model projections that compensate for the amount-intensity discrepancy, and to develop an approach to transform observed rainstorms into future simulations to account for those changes. The remainder of this paper is organized as follows. In Section \ref{section:ModelAndObs}, we describe the high-resolution regional climate model runs and the radar-based observational data products used in the study.  In Section \ref{section:RainstormAnalysis1}, we develop an algorithm for identifying and tracking individual rainstorms, and in Section \ref{section:RainstormAnalysis2}, we develop metrics for analyzing spatio-temporal precipitation patterns. In Section \ref{section:results}, we compare precipitation in observations and present-day and future climate projections, and in Section \ref{section:simulation}, we develop a simulation method to generate future precipitation simulations that combines information from models and observations. Finally, in Section \ref{section:discussion}, we discuss the implications of these results.

\section{Climate Model Output and Radar-based Observational Data} \label{section:ModelAndObs}

\begin{figure}
\begin{center}
\includegraphics[scale=0.54]{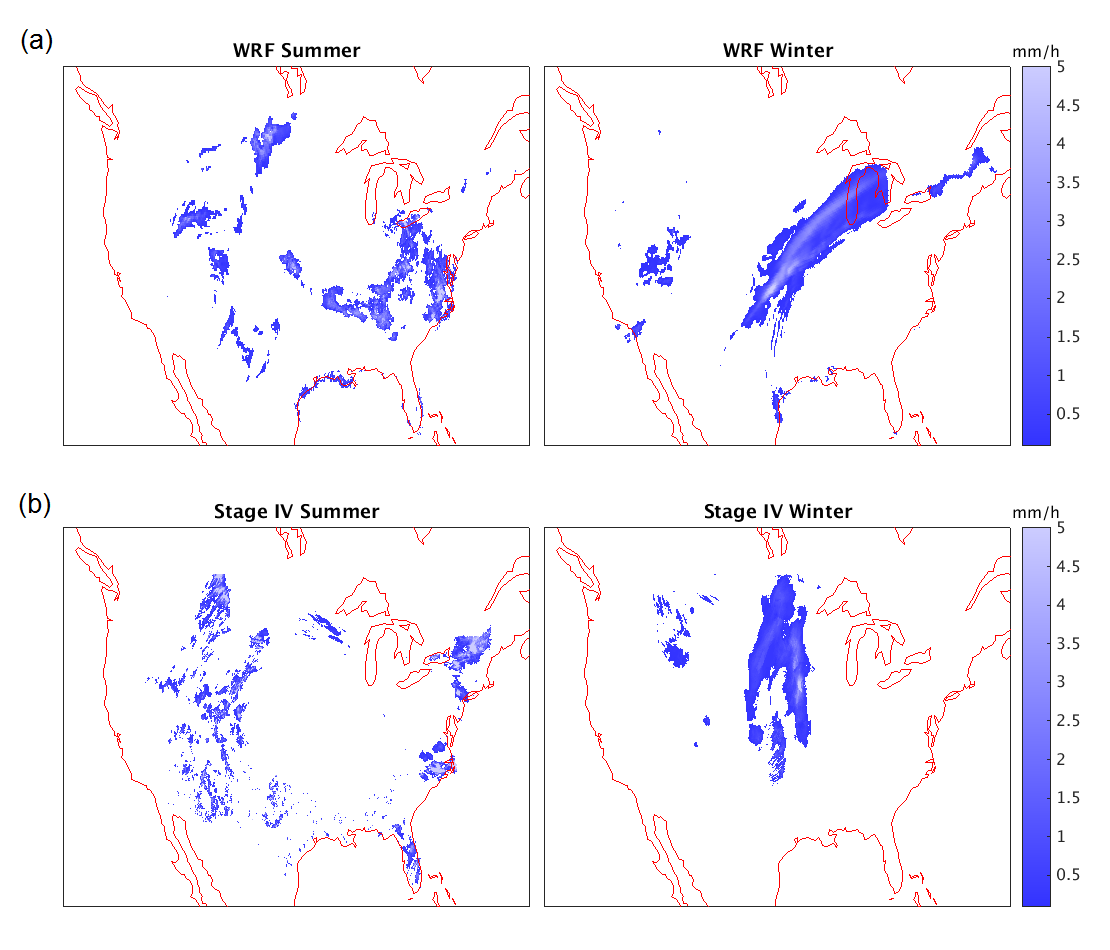}
\end{center}
\caption{Example precipitation patterns in a single time step from (a) WRF and (b) Stage IV data, for summer  (left) and winter (right) seasons. Precipitation events in model and observations show similar morphological characteristics: summer precipitation appearing as broad regions of small-scale convection and winter precipitation as coherent larger-scale phenomena.}
\label{figure:illustration}
\end{figure}

As discussed above, model output used to study rainstorm characteristics must be at high spatial resolution.  In this study 
we use high-resolution dynamically downscaled model runs over the continental U.S.\ region, with a constant-spacing grid of 12 $\times$ 12 km, described by \citet{wang2015high}. These runs use the Weather Forecasting and Research (WRF) model \citep{skamarock2008time} as the high-resolution regional climate model, nudged by a coarser simulation from the Community Climate System Model 4 (CCSM4) \citep{gent2011community} of the business-as-usual (RCP 8.5) scenario. (The CCSM4 run is an ensemble member from the CMIP5 archive; see \cite{meinshausen2011rcp} and \cite{WCRP2015CMIP5} for further information.) {\color{black} In the WRF simulation we apply spectral nudging, which is conducted in the interior as well as at the lateral boundaries, on horizontal winds, temperature, and geopotential height above 850 hPa. } %Reviewer 1 Specific 4
Because high-resolution runs are computationally demanding, WRF was run only for two 10-year segments of the scenario, which we term the ``baseline'' (1995-2004) and ``future'' (2085-2094) time periods. In this analysis we separately analyze summertime (June, July, August) and wintertime (December, January, February) precipitation because precipitation characteristics differ by season. U.S.\ precipitation is predominantly convective in summer and large-scale in winter (Figure \ref{figure:illustration}).

Both CCSM4 and WRF are widely-used models for atmospheric and climate science. WRF has been extensively used for mesoscale convection studies, for dynamical downscaling of climate model projections \citep{lo2008assessment,bukovsky2009precipitation,wang2015model}, and as a forecasting model for numerical weather prediction \citep[e.g.][]{jankov2005impact,davis2006object2,clark2009comparison,skamarock2008time}.  Validation studies addressing these different contexts include \citet{ma2015dynamic}, who showed in a model-observation comparison that downscaling CCSM4 output using WRF improved the fidelity of modeled summer precipitation over China on both seasonal and sub-seasonal timescales. For forecasting, \citet{davis2006object2} studied intense and long-lived storms in 22-km resolution WRF forecasts over the U.S.\ and found that the model captured storm initialization reasonably well but showed some apparent biases in storm size, intensity, and duration. 

\begin{figure}
\begin{center}
\includegraphics[scale=0.49]{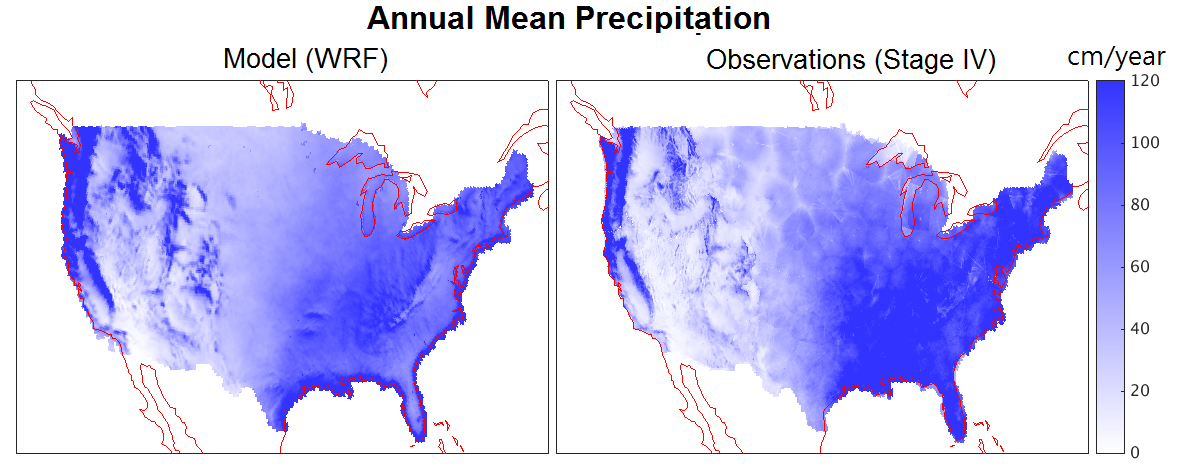}
\end{center}
\caption{Comparison of time-averaged precipitation in the baseline model run and in Stage IV observational data. (We use the 24 hourly Stage IV dataset here, to provide full coverage over the West Coast.) While mean precipitation over the study area is similar in model and observations (77 vs.\ 72 cm/yr, respectively), the model shows strong spatial discrepancies, underestimating precipitation in the East and overestimating it in the West.  Note mosaicing effects in the observations.}
\label{figure:SpatialAndHistogram}
\end{figure}

We compare WRF/CCSM4 model output to an observational data product of similar spatial and temporal resolution: the NCEP Stage IV analysis (``Stage IV data''  hereafter), which is based on combined radar and gauge data \citep{lin2005ncep,prat2015evaluation}. The Stage IV dataset provides hourly and 24-hourly precipitation at 4 km resolution (constant-spacing $4\times 4$ km grid cells) over the contiguous United Stages from 2002 to the present. (To match our model output, we use ten years of data from 2002 to 2011, and aggregate the 1-hourly data to 3-hourly.) 
Stage IV data are produced by `mosaicing' data from different regions %estimated based on multi-sensor precipitation analyses 
into a unified gridded dataset. This mosaicing leaves artifacts in the spatial pattern of average precipitation: the range edges of individual radar stations are visible by eye in Figure \ref{figure:SpatialAndHistogram}.  These artifacts do not compromise our analysis of rainstorm frequency, duration, and size, {\color{black}as
mosaicing effects do not affect temporal evolution. Stage IV data have been shown to be temporally well-correlated with high-quality measurements from individual gauges. (See e.g., \citealp{sapiano2009intercomparison,prat2015evaluation}.)}

\begin{figure}
\begin{center}
\includegraphics[scale=0.65]{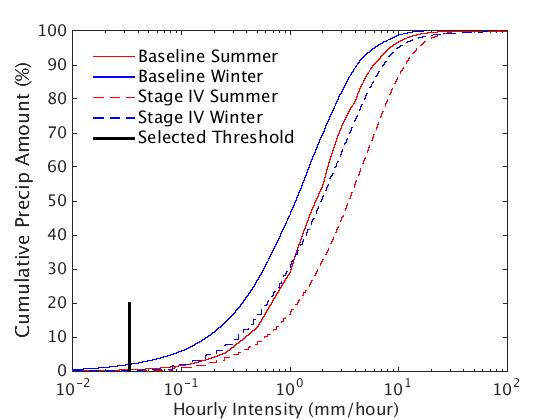}
\end{center}
\caption{Cumulative distributions of precipitation vs.\ intensity in the baseline model output (solid) and in observations (dashed), for the entire study area eastward of $114^\circ$ W (i.e. excluding locations where observations are problematic). Summer is shown in red and winter in blue for both cases. Data used are 3-hourly precipitation for all individual and model grid cells (i.e.\ no spatial or temporal aggregation), with units restated as hourly precipitation for clarity. High-intensity precipitation contributes more heavily to cumulative precipitation in observations than in model output, in both seasons. Differences here imply that rainfall intensity is $\sim 50$\% lower in model output than in observations.
% XX could delete it or move to text if needed
 This difference is unlikely to be a sampling artifact of spatial resolution since both model and observation grid cell sizes (144 and 16 km$^2$) are considerably smaller than typical storm sizes.
 Black line marks our cutoff intensity threshold for the analysis, 0.33 mm/hour. This cutoff negligibly affects the analysis; 
even in the worst case (model winter) it excludes only 2\% of total precipitation. %{\color{red}We did not include the observational data from the region west of $114^\circ$ W to remove the area where the Stage IV data are problematic.  The model output in the same region has been also excluded for consistency.}
}
\label{figure:lost_amount}
\end{figure}

%  possibly because the WRF model has difficulty in reproducing storms that originate outside the high-resolution area of the dynamically downscaled model. 
Comparing the present-day WRF run with Stage IV data shows that the model captures seasonal mean precipitation amounts well (Tables \ref{table:SummerDisc} and \ref{table:WinterDisc}, the first row), but with a large bias in rainfall intensity. Rain rates in precipitating grid cells in WRF output are $\sim$50\% lower than in observations (Figure \ref{figure:lost_amount}). The bias is consistent across two orders of magnitude in intensity, nearly identical in both summer and winter, and unlikely to be a sampling artifact, since both model and observation grid cell sizes (144 and 16 km$^2$) are well below typical areal sizes of precipitating events. Because the model matches observed total rainfall, the too-low intensity in individual rainstorms must be compensated by some other bias in precipitation characteristics:
model precipitation events may be more frequent, more numerous, larger in size, or any combination of these factors.  (\citet{davis2006object2} has noted a bias in WRF toward too-large rainstorm events, though in a study of only the most severe storms.)  Understanding model-observation biases therefore becomes an analogous problem to understanding differences in present and future model projections. Both cases involve amount-intensity discrepancies, which imply some change in spatio-temporal precipitation properties that can be understood only by 
%the physical mechanisms underlying an amount-intensity discrepancy requires 
identifying and tracking individual rainstorms. % and analyzing their spatio-temporal properties. 

Intensity differences do introduce one minor complication in analysis. It is typical in statistical analyses of precipitation to cut off all data with precipitation intensity below a particular threshold, to remove spurious light precipitation and make analysis tractable. In our case the intensity distributions for model and observation are different, suggesting that different cutoffs might be warranted.  
For simplicity, we apply the same cutoff of 
 $0.033$ mm/hour (black line in Figure \ref{figure:lost_amount}) to both model and observations. This choice should negligibly affect the overall results, since even in the worst case (model wintertime), the cutoff excludes less than 2\% of seasonal precipitation.

\section{Identifying and Tracking Individual Rainstorms} \label{section:RainstormAnalysis1}

As discussed previously, existing algorithms for finding precipitation events are appropriate only for severe storms. To decompose  the entire precipitation field into a set of events, we develop here a more general approach. We define a ``rainstorm'' for this purpose as a set of precipitating grid cells that are close in location and that move together retaining morphological consistency. As in \citet{davis2006object1}, we use only the precipitation field itself, allowing ready comparison to observations. 
(More complex definitions might also draw on information about the meteorological context.)  

Defining rainstorm events in spatio-temporal data involves two tasks
\begin{itemize}
\item[A.] Rainstorm identification: Divide the precipitation field at each time step into individual storms% (Subsection \ref{subsection:StormIdentification})
\item[B.] Rainstorm tracking: Build rainstorm events evolving over time by tracking identified storms across consecutive time steps% (Subsection  \ref{subsection:StormTracking})
\end{itemize}
We describe each step below, and provide more detail on algorithms in Section S1 in Supplemental Material. 

\subsection{Identifying rainstorms at a single time step} \label{subsection:StormIdentification}
\begin{figure}
\begin{center}
\includegraphics[scale=0.25]{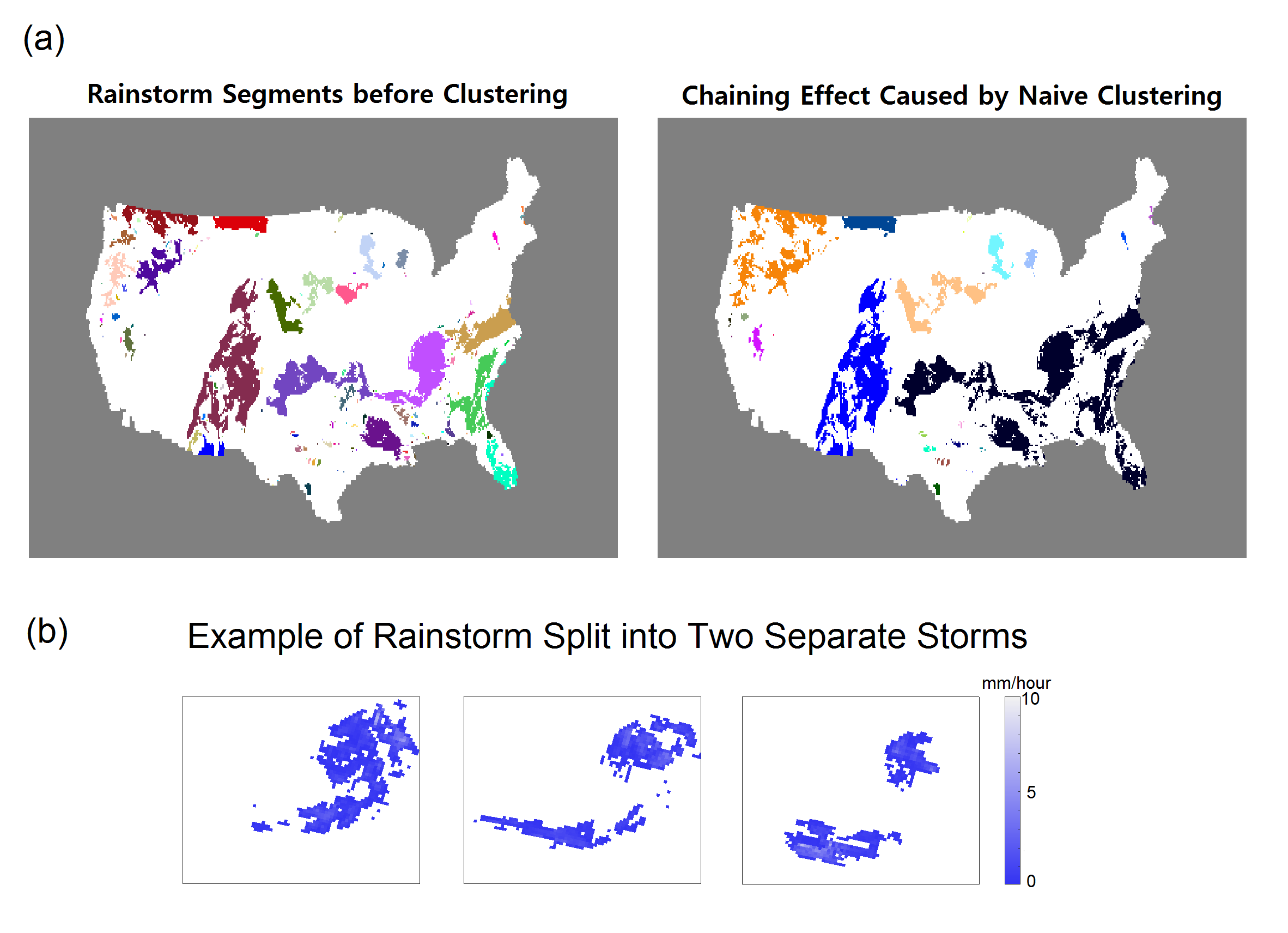}
\end{center}
\caption{Challenges in rainstorm identification and tracking. a) shows two imperfect clustering strategies, illustrated by a representative timestep in model output. Left: simple linking of contiguous precipitation regions produces too many identified events. Right: a naive almost-connected-component clustering algorithm produces too few identified events, due to the chaining effect. b) shows an example of a rainstorm split into two different rainstorms over time. Our tracking algorithm must be able to track the sub-storms and label them as the same rainstorm event.}
\label{figure:challenges}
\end{figure}

We first identify rainstorms from the precipitation field at each time step $t$. The simplest approach would be to treat each contiguous region of grid cells with positive precipitation exceeding some threshold criterion as an individual rainstorm \citep{guinard2015projected}. This algorithm however identifies excessive individual rainstorm events that are in reality meteorologically related: a single storm system can often have multiple separate precipitating regions (Figure \ref{figure:challenges}a, left panel).  Rainstorm identification is properly a clustering problem that requires grouping related but not necessarily contiguous phenomena.
We group nearby regions into ``rainstorm segments'' using a technique termed ``almost-connected component labeling'' \citep{eddins2010almost,murthy2015automated}. The basic idea is to identify precipitating regions 
close enough they would connect given some prescribed dilation of individual regions. This procedure provides a natural way of grouping regions based on their proximity and morphological features and has been previously used for rainstorms \citep{baldwin2005development}. Unlike other traditional clustering algorithms such as k-means clustering \citep{hartigan1979algorithm}, the approach does not require a pre-specified number of clusters and hence enables quick and automatic clustering. 
 
\begin{figure}
\begin{center}
\includegraphics[scale=0.18]{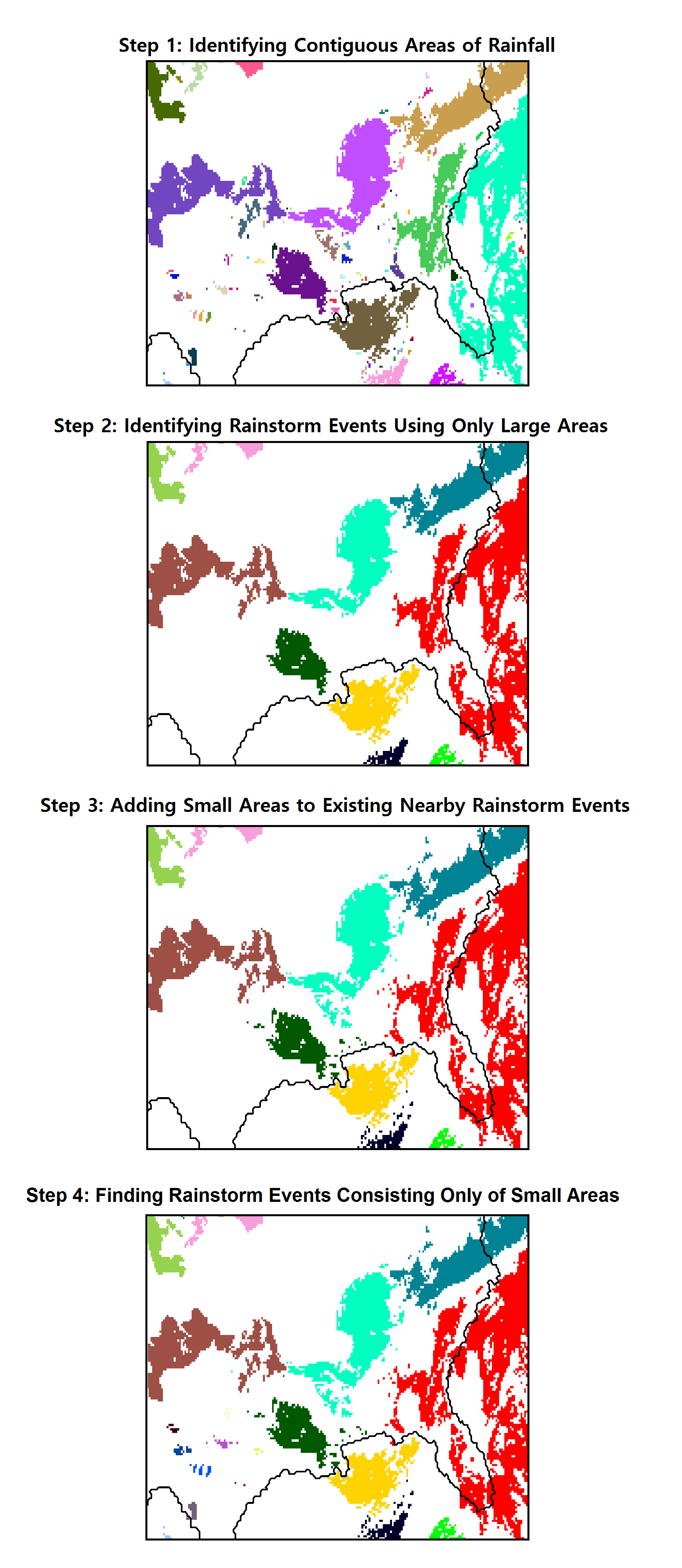}
\end{center}
\caption{Illustration of our rainstorm identification algorithm with representative summertime model output. In Step 1, we identify all contiguous precipitation areas. In Step 2, we apply almost-connected-component labeling for only the large areas. In Step 3, we add small areas to the existing nearby rainstorm events if they are close enough to any existing ones. In Step 4, we form rainstorm events that consist of only the remaining small areas. This approach prevents the chaining effect shown in Figure \ref{figure:challenges}a.}
\label{figure:identification}
\end{figure}

Almost-connected component labeling does often suffer from a ``chaining effect'' that produces overly large rainstorm segments: a small number of precipitating grid cells located between two large rain regions can cause awkward linkage between them (Figure \ref{figure:challenges}a, right panel). To avoid this, we use a four stage  procedure based on mathematical morphology that treats ``large'' and ``small'' regions separately.  
(A similar approach was used on radar reflectivity datasets by \cite{han20093d}.) 
We first identify contiguous regions of precipitation (Figure \ref{figure:identification}, step 1; our threshold is $>0.1$ mm/3 hour). 
In the second stage, we form segments using only the ``large'' regions (Figure \ref{figure:identification}, step 2). In the third stage we assign small regions to the closest existing segments from the previous stage if they are close to the existing segments (Figure \ref{figure:identification} step 3). In the fourth stage we form segments with the remaining small regions (Figure \ref{figure:identification} step 4). %See Section XX?? in the Supplementary Document for a more detailed description of the algorithm. XX ... presumably the reference to the appendix is done up front now

\subsection{Tracking rainstorms over different time steps} \label{subsection:StormTracking}

Once the rainstorm segments for all time points are identified, we link them in different time steps to form rainstorm events evolving over time and space. This task is complicated by the fact that rainstorms  often split into multiple segments that drift in different directions. (See Figure \ref{figure:challenges}b for illustration.) Most existing algorithms are not designed to  handle this possibility. Our new algorithm is partly inspired by the work of \cite{hodges1994general} and \cite{morel2002climatology1}, but extends their work to more efficiently identify segments originating from the same rainstorm and track their movement over time.

\begin{figure}
\begin{center}
\includegraphics[scale=0.23]{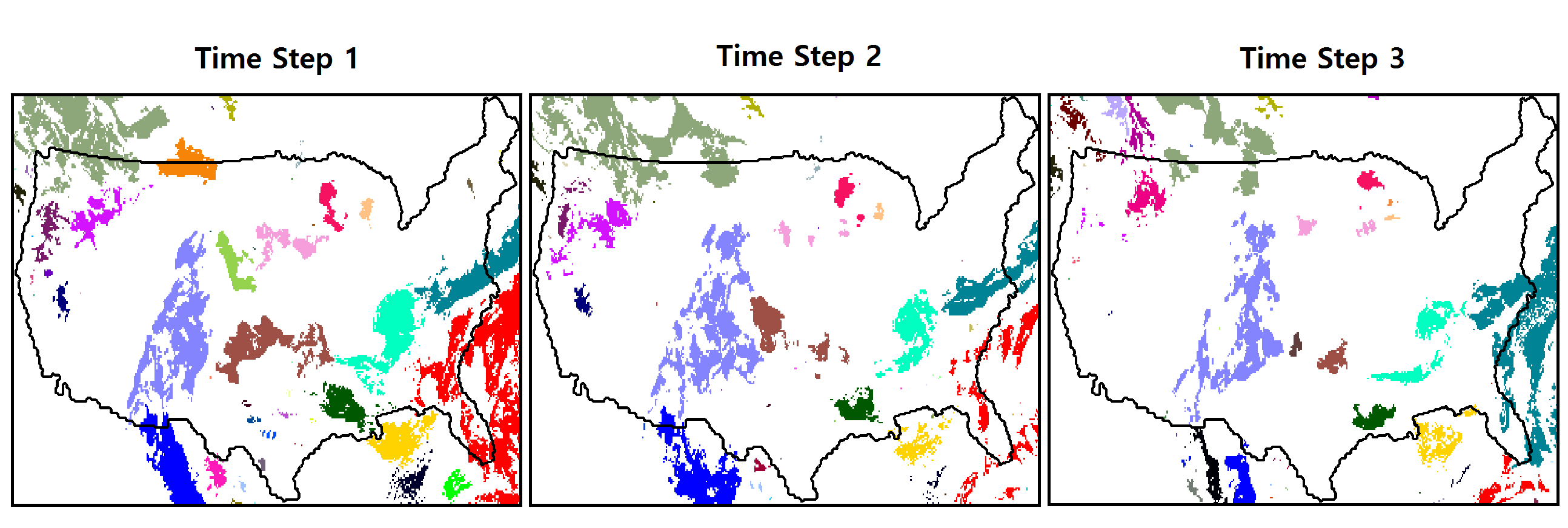}
\end{center}
\caption{Rainstorm objects constructed by our tracking algorithm in three consecutive example time steps in model output, distinguished by different colors. Our algorithm can efficiently track multiple events simultaneously and represent various storm merging and splitting situations. This example contains a storm merger in the northwest (orange and dark-green storms combine) and a storm split in the southeast (light-green storm splits into multiple segments.}

\label{figure:tracking}
\end{figure}

The algorithm works sequentially in time. For each time step from $t = 1$ to $T$,  we assign each rainstorm segment to one of the existing rainstorm events based on the two criteria:  (i) the shapes of linked rainstorms in two consecutive time steps are morphologically similar enough to be considered as the same rainstorm, and (ii) the rainstorm location and the movement direction do not change too abruptly over time. If we cannot find any existing rainstorm events that meet these criteria for a rainstorm segment, we initialize a new rainstorm event starting with that segment. On the other hand, if more than two events satisfy the criteria, we assign the rainstorm segment to the largest event. Since the multiple rainstorm segments at the same time step can be assigned to the same rainstorm event, our algorithm can incorporate situations where a rainstorm splits into multiple segments. %Section ?? in the Supplementary Document provides a more detailed description of the algorithm. 
We denote the resulting rainstorm events $S_1,\dots,S_N$. 
%for the baseline scenario by $S_1^1,\dots,S_{N_1}^1$  and for the future scenario by $S_1^0\dots,S_{N_0}^0$, where $N_0$ and $N_1$ are the number of rainstorm objects for each scenario. Similarly we denote the rainstorm objects from the observational data by $Z_1^0,\dots,Z_{M_0}^0$.  
Figure \ref{figure:tracking} shows example time steps for rainstorm tracking results from our algorithm, which appears to simultaneously track different rainstorms with various morphological features reasonably well. %Moreover, the algorithm can identify the rainstorm segments originated from the same rainstorm and include them in the same rainstorm event.

% XX shortened title
\section{Describing Rainstorm Characteristics} \label{section:RainstormAnalysis2}

% XX this is stating that you'll find some results, but in this section you're just stating metrics
% XX just say that, you can do results later
%We characterize individual rainstorms found by the algorithm described in the previous section and analyze the precipitation patterns in the model runs and the observational data based on those characteristics. 
%Using the identified rainstorm events we analyze the model-observation discrepancy and the projected future changes through various rainstorm characteristics.

Once rainstorms are identified and tracked, the goal is to  
quantitatively describe them.   
There is little precedent in the literature for this task;
most previous rainstorm studies focus only on analyzing storm trajectories or tracks \citep{hodges1994general,morel2002climatology2}. % XX no need to beat up on them more 
%perhaps due to the limitation of the existing rainstorm identification and tracking algorithms explained above. 
We therefore develop a set of metrics for characterizing storms, and follow a four-stage process for  
describing spatio-temporal precipitation patterns and comparing those patterns between different climate states (or between model output and observations):			
% XX good phrase, save
%fully understand the model-observation discrepancy and the future projections in terms rainstorm characteristics. 
\begin{itemize}
\item[A.] Compute metrics for individual rainstorms: duration, size, mean intensity and central location.% (Subsection \ref{subsection:metrics})
\item[B.] Extend those metrics to apply to aggregate precipitation. That is, we decompose the total precipitation amount into the product of: average intensity, a size factor, a duration factor, and the number of rainstorms. % (Subsection \ref{section:factorization}).
\item[C.] Estimate the geographic variation of those aggregate metrics. That is, we find spatial patterns of rainstorm properties.  %(Subsection \ref{section:SpatialAnalysis})
\item[D.] Compare precipitation patterns across model runs or across models and observations. % : Using the computed factors and the spatial patterns we compare different precipitation patterns from model runs and observational data in terms of rainstorm characteristics (Subsection \ref{section:ComparisonMethod}). 
\end{itemize}
We describe the algorithms for each of these step below.

\subsection{Metrics for individual rainstorms} \label{subsection:metrics}

%Using the rainstorm events found by the above algorithm we can characterize the model runs and observational data in terms of rainstorm properties. To this end 
We characterize each individual rainstorm event $S_i$ with four metrics: duration, size, mean intensity, and central location. 
For completeness, we describe below how each metric is computed. Note that size, location, and amount metrics are not scalars but vectors or matrices over the lifetime of each storm.

%\item \textit{Duration} (the length of time that the rainstorm event lasted, in hours) 
%\item \textit{Size} (a timeseries of spatial areas covered by the rainstorm event during its lifetime, a vector with units of km$^2$)
%\item \textit{Central location} (a timeseries of the latitude and longitude of the center of the rainstorm event) 
%\item \textit{Total precipitation amount}: (a timeseries of total precipitation in the rainstorm event for each 3-hour timestep)

\textit{\textbf{Duration}}. The storm identifying algorithm gives us the beginning and ending timesteps of the lifetime of storm $S_i$ ($b(S_i)$ and $e(S_i)$, respectively). Since all analysis involves 3-hour timesteps, the rainstorm duration is $l(S_i)\times$3 hours, where $l(S_i)=e(S_i) -b(S_i)+1$. 

\textit{\textbf{Size}}. Rainstorm size is a vector of length $l(S_i)$ timesteps. The storm identification algorithm identifies a number $s(S_i,t)$ of grid cells as part of the storm at each timestep $t$. (We do not need to estimate fractional coverage of individual grid cells when using a grid as fine as 12 $\times$ 12 km or 4 $\times$ 4 km.)  The size at each timestep is the $s(S_i,t)\times$144 km$^2$ for the model output and $s(S_i,t)\times$16 km$^2$ for the observational data.

\textit{\textbf{Mean intensity}}. Also a time series of length $l(S_i)$. At each time step $t$ when the rainstorm $S_i$ exists over the contiguous U.S.\, we compute its average precipitation amount $a(S_i,t)$ by taking the average of the precipitation intensity over all grid cells identified with the storm, i.e.\ $a(S_i,t)=\sum_{k=1}^{s(S_i,t)} v_{i,t,k} / s(S_i,t)$, where $v_{i,t,k}$ $\left(k=1,\dots,s(S_i,t)\right)$ is the precipitation intensity at each grid cell location of $S_i$ at time $t$.

\textit{\textbf{Central location}}. The central location of each rainstorm event $S_i$ is a $l(S_i)\times2$ matrix, where each row $c(S_i,t)$ is the the center of gravity weighted by the precipitation amount in each grid cell. This location measure is invariant to rotation and translation of the rainstorm event on the surface of the globe.  (See Section S2 in the Supplemental Material.)

Application of this procedure does still result, in both model output and observations, in large numbers of small rainstorms with negligible precipitation, many with very short lifetimes. These tiny storms would not affect estimates of average storm intensity and size, which are weighted by precipitation amount and event size, but if included could dominate estimates of number of storms and their mean duration, reducing the utility of these factors. Because our goal is to provide information about events that contribute significantly to overall rainfall,  we therefore exclude in all subsequent analyses those rainstorms with the lowest individual precipitation amounts. That is, we remove those small events in the tail of the cumulative distribution whose values add up to 0.1\% of total precipitation.
 
%\end{itemize}
   
%--- EJM edited this section to here 11/21 --------------------
% --- XX remove the list-within-a-list below -------------------
 
\subsection{Factorizing total precipitation} \label{section:factorization}

Using the computed metrics for each rainstorm, we can summarize the spatio-temporal aspects of precipitation patterns over a specified region by decomposing the total precipitation amount into the following four factors:
\begin{align*}
\mbox{Total Amount} =  \mbox{Average Intensity} \times \mbox{Size Factor}\times \mbox{Duration Factor} \times \mbox{Number of Rainstorms}.
\end{align*}

%{\color{red} The approach described here allows us to decompose the total precipitation into easily interpretable factors, but it is completely uninformative about subregional differences within the analysis domain. This motivates the spatial analysis described in Section .} 

Our definition of these factors (see below) is chosen to ensure that a given fractional change in any factor leads to the same fractional change in total precipitation amount. That is, the factors allow us to quantitatively compare the effects of changes in different rainstorm properties. In Section 5, we use these factors to compare the model baseline period to Stage IV data to evaluate model-observation discrepancies in rainstorm characteristics.  We also use the same approach to compare the model baseline and future periods to identify changes in spatio-temporal precipitation patterns and quantify how they contribute to the total precipitation amount change. Factorization results are presented in Tables 1--6. Such an analysis necessarily involves aggregating events over some reasonably large region; to identify subregional differences we also conduct a spatial analysis described in subsection \ref{section:SpatialAnalysis} below. 

Each factor is defined as follows:

\textit{\textbf{Average intensity (mm/hour)}}. The average precipitation intensity is the size-weighted average of all rainstorm intensities. Since the length of each time step is 3 hours, the hourly intensity is computed as $\frac{1}{3} \times \sum_{i=1}^{N} \sum_{t=b(S_i)}^{e(S_i)} a(S_i,t) \times s(S_i,t) / \sum_{i=1}^{N} \sum_{t=b(S_i)}^{ e(S_I)} s(S_i,t)$.

\textit{\textbf{Size factor (km$^2$)}}. Size factor is the average storm size per each rainstorm event at each time step, computed as $\mbox{(the area of each grid cell)} \times \sum_{i=1}^{N}\sum_{t=b(S_i)}^{e(S_i)} s(S_i,t) / \sum_{i=1}^{N} l(S_i)$, where the area of each grid cell is $144 \mbox{ km}^2$ for the model output and $16 \mbox{ km}^2$ for the observational data.

\textit{\textbf{Duration factor (hour/rainstorm)}}. Duration factor is the mean duration per each rainstorm event. The factor is computed as $3 \mbox{ hours}\times\sum_{i=1}^{N}  l(S_i)  / N$, because the length of each time step is 3 hours.

\textit{\textbf{Number of Rainstorms}}. The number of rainstorms is simply given by $N$.

\subsection{Spatial analysis} \label{section:SpatialAnalysis}

%We can also use the computed metrics for rainstorm events for spatial analysis of rainstorm characteristics. 
%We can find and visualize the spatial distribution of the metrics over the area of interest and compare the spatial patterns between the baseline and the future runs to understand the projected changes, or the baseline run and the observational data to study the model-observation discrepancy. 
We can also use the computed metrics to find and visualize the spatial distribution of rainstorm characteristics rather than aggregating regionally. Precipitation intensity can be simply computed at individual grid cells without making use of rainstorm identification, but size, duration, and number require rainstorm identification and some statistical method for assigning a value to an individual grid cell. We use for this purpose a non-parametric kernel estimator, a standard procedure in spatial point process analysis \citep[see, e.g.][Chapter 21]{gelfand2010handbook}. 
The procedure is most simply illustrated for estimating 
the expected number of rainstorms $n$ originating at a given spatial location $\mathbf{s}$, as 
\begin{equation*}
n(\mathbf{s})=\sum_{i=1}^{N} k(c(S_i,1),\mathbf{s}).
\end{equation*}
Here $k(\mathbf{s}_1,\mathbf{s}_2)$ is a kernel function that satisfies $\sum_{\mathbf{s}_2 \in \mathcal{S}} k(\mathbf{s}_1,\mathbf{s}_2)=1$ and smoothly decays as the distance between $\mathbf{s}_1$ and $\mathbf{s}_2$ increases, and $\mathcal{S}$ is a set of all grid cell locations in the contiguous United States. (See Section S3 in the Supplemental Material for details about the choice of the kernel function.) 
For rainstorm size and duration, we use a similar kernel estimator but now compute the spatially-weighted average value of the rainstorm metric, with the spatial weighting given by the kernal function. 
For further details see Section S4 in the Supplemental Material. {\color{black}(Note that the regional factors produced by this analysis will not sum exactly to the total changes in seasonal mean precipitation, especially when seasonal mean precipitation is determined as an unsmoothed quantity computed at each grid cell.)}

\subsection{Comparing Rainstorm Characteristics in Two Precipitation Patterns} \label{section:ComparisonMethod}

The approaches described in Section 4\ref{subsection:metrics}--\ref{section:SpatialAnalysis} above provide a framework to compare any two spatio-temporal precipitation datasets and to identify similarities and differences in rainstorm characteristics. In the analysis below, we decompose total precipitation amounts for our different model and observational datasets into their component factors using the methods of Section 4\ref{section:factorization}, and to identify differences, we compute the ratios of the factors. This analysis helps us to understand, in an average sense, which characteristics of rainstorms drive differences in spatio-temporal precipitation. To identify geographic variations in the various rainstorm characteristics, we also use the spatial analysis 
of Section 4\ref{section:SpatialAnalysis} to map individual rainstorm metrics across the study area, and again compare the different datasets by taking ratios. 
To quantify the statistical uncertainty in estimating the ratios of factors, we construct 95\% confidence intervals by bootstrap sampling (shown in last column of Tables \ref{table:SummerChange}--\ref{table:WinterChangeDry}). See Section S5 in the Supplemental Material for details of uncertainty analysis.

\section{Results} \label{section:results}

The methods described above allow us to compare rainstorm characteristics both between model output and observations and between model output for present-day and future climate states. In both cases we know \textit{a priori} that there must be differences in spatio-temporal precipitation characteristics.  %{\color{red} The results that we present in this section are, of course, specific to the model and data used in the analysis here.  However, the developed approach in the previous section can readily be applicable to other model runs and observational datasets for similar analysis. } 

\subsection{Model-observation comparison} 
\begin{figure}
\begin{center}
\includegraphics[scale=0.26]{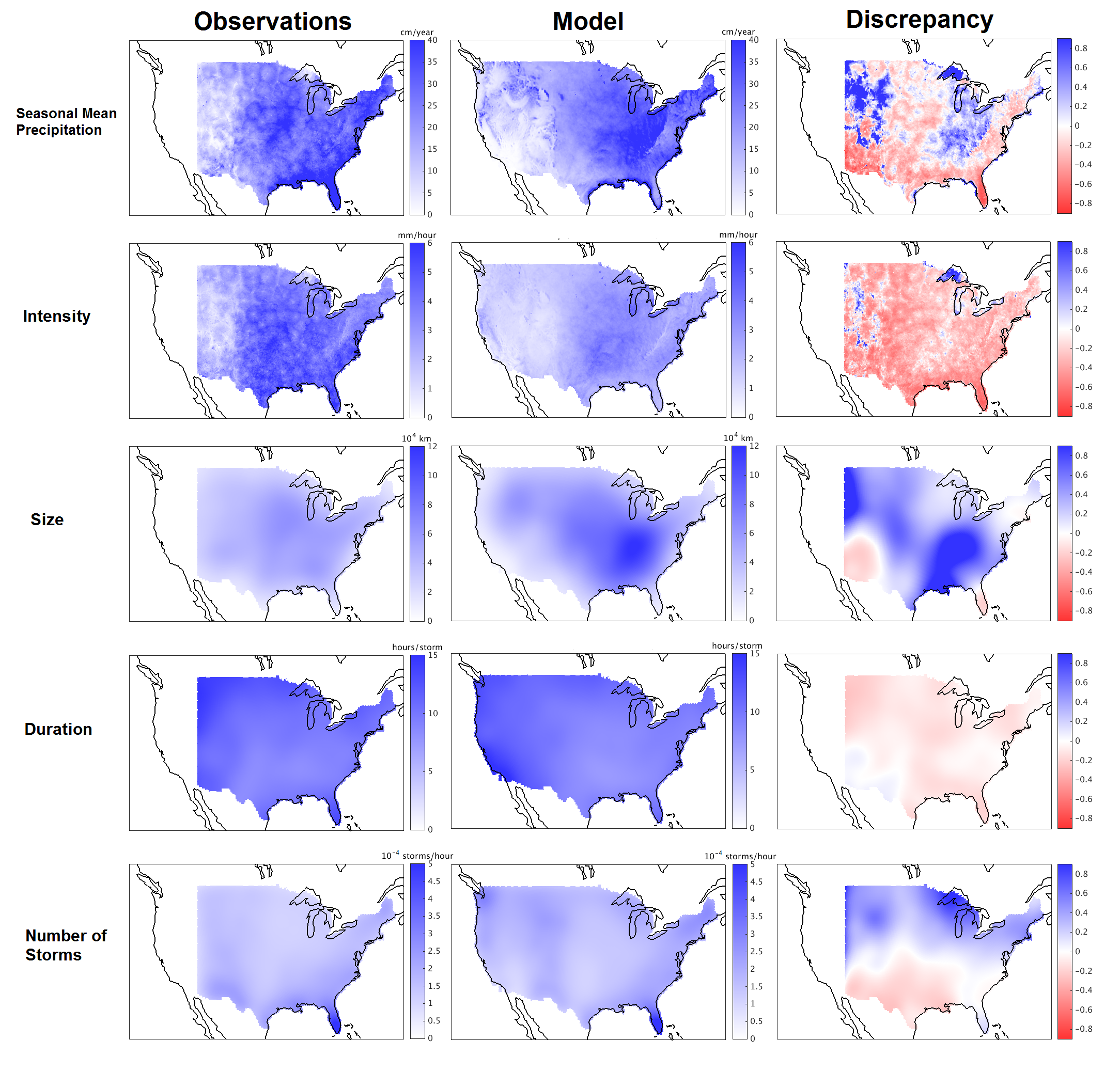}
\end{center}
\caption{Model-observation comparison for the summer season. First and second columns show precipitation and computed storm characteristics for observations and model, respectively; third column shows their fractional difference (model bias). Note that seasonal mean precipitation and intensity (rows 1--2) are computed for individual grid cells while size, duration, and number of storms (row 3--5) are computed with kernel smoothing as described in Section 4c. Model rainstorm events are too large and too weak. Rainstorm initialization number shows a geographic pattern of bias, with the model producing too many rainstorms in the north and too few in the south. Despite these biases the U.S.-average model total seasonal precipitation matches observations well. (See also Table 1.)}
\label{figure:SummerDisc}
\end{figure}

\begin{figure}
\begin{center}
\includegraphics[scale=0.26]{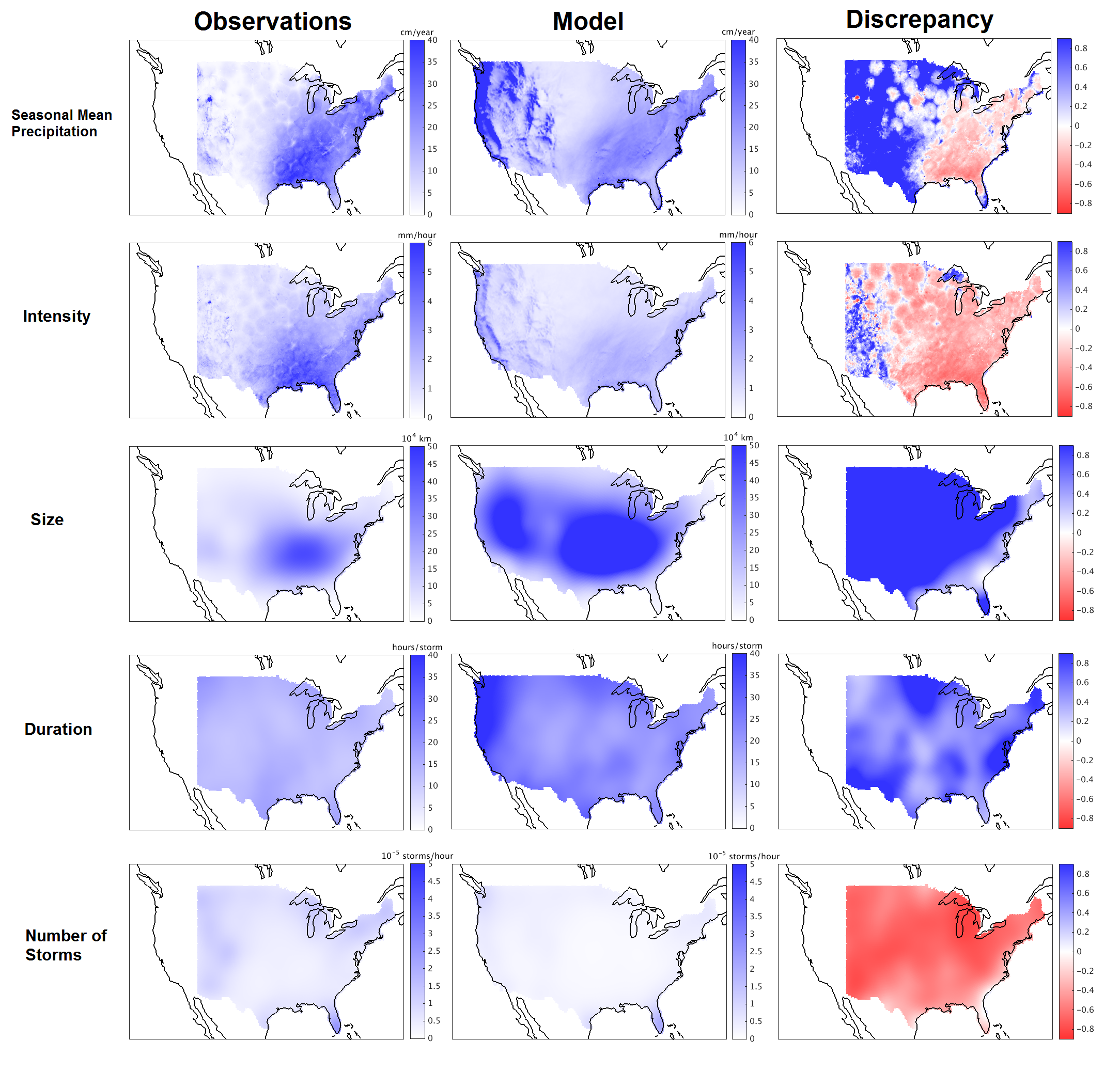}
\end{center}
\caption{As in Figure \ref{figure:SummerDisc} but for the winter season.  As in summer, model rainstorm events are generally too large and too weak, but now also less numerous and too long in duration. Size and intensity biases show an east-west gradation that results in east-west differences in model bias in total precipitation, too dry in the east and too wet in the west. Despite these biases the model U.S.-average total seasonal precipitation matches observations well. (See also Table 2.)}
\label{figure:WinterDisc}
\end{figure}

We find that WRF output and Stage IV data show substantial differences in rainstorm characteristics, but with different discrepancies in summer (Figure \ref{figure:SummerDisc} and Table \ref{table:SummerDisc}) and winter (Figure \ref{figure:WinterDisc} and Table \ref{table:WinterDisc}). In both seasons, we find that intensity in modeled precipitation events is considerably too low (as was evident even without rainstorm identification) even though U.S.\-average total precpitation is similar in model and observations. 
In summer, the compensating factor that resolves this discrepancy is predominantly size.  WRF summer rainstorms are roughly 30\% too weak but also 50\% too large (Table 1). The size bias is similar to that found by \citet{davis2006object2} for severe summmer storms in a single-year WRF simulation. 
While \citet{davis2006object2} reports that the duration of severe storms is too long, we find no overall bias in summer rainstorm duration in WRF. In our model output, the total number of summer storms across the study area is similar to that in observations, but with geographic patterns of bias: too many storms in the north and too few in the south (Figure \ref{figure:SummerDisc}).

In winter, when U.S.\ precipitation is predominantly large-scale rather than convective, model biases are both qualitatively and quantitatively different than in summer. 
The intensity and size effects seen in summer become still larger in winter, and other factors become discrepant as well.
WRF winter rainstorms are substantially weaker, larger, fewer, and longer than those in Stage IV data. 
 (Figure \ref{figure:WinterDisc} and Table \ref{table:WinterDisc}). Winter storms are half as intense in the model as in observations and more than twice as large, biases which roughly cancel.
The number of winter storms is also half that of observations, and their duration twice as long. Changes in storm number are nearly uniform across the study area,
% XX check this - figure is saturated
but other biases do not cancel locally. They do approximately cancel across the study area, so that while local precipitation amounts can be discrepant from observations, model and observations again are in reasonable agreement for total U.S.\ average precipitation.

\subsection{Model present-future comparison} 
While the model-observation comparison is sobering, WRF model runs offer a self-consistent response to changed climate conditions {\color{black} that is useful to study. 
The simulations must involve some compensating change that reconciles the diverging increases in precipitation amount and intensity. Different models may in fact find differing solutions to these constraints, but the way to gain insight into potential future changes in precipitation characteristics is to carefully examine model responses.}

\begin{figure}
\begin{center}
\includegraphics[scale=0.26]{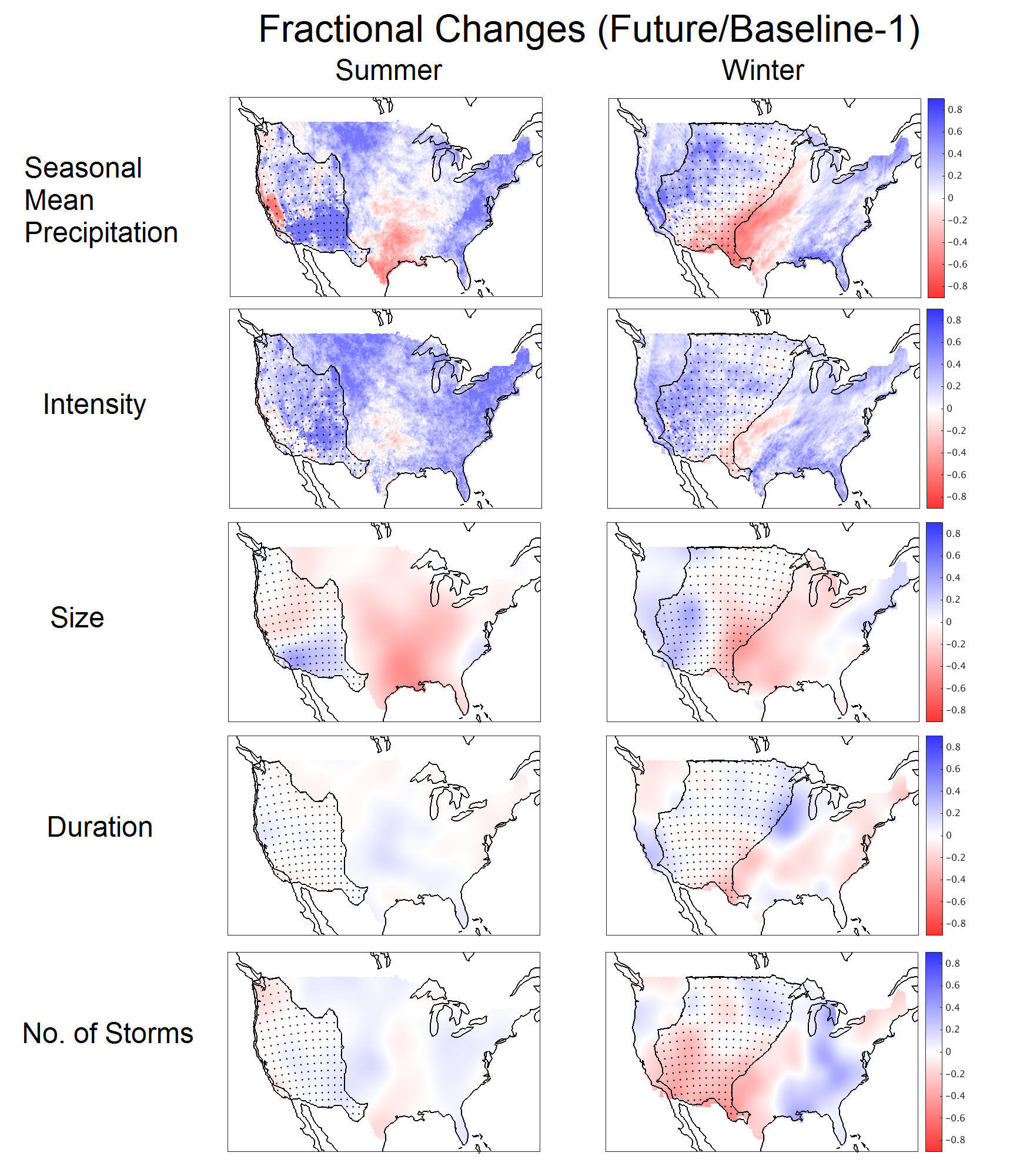}
\end{center}
\caption{Fractional changes in total precipitation and rainstorm characteristics between model baseline and future runs, for both summer and winter. Dotted areas roughly correspond to dry regions with precipitation less than 10 cm/season. As in Figures \ref{figure:SummerDisc} and \ref{figure:WinterDisc}, changes in total precipitation and intensity (rows 1--2) are computed at individual grid cells and those in other factors are kernel-smoothed. In both seasons, total precipitation largely increases, but with areas of drying in the Southwest, extending into the Great Plains in winter. Intensity increases slightly more uniformly than does total precipitation.  Rainstorm size decreases substantially in most regions; duration and number of storms do not show clear changes. Overall spatial patterns of changes are more complex in winter than in summer.}
%The area with decreased annual mean is much larger than the summer seasons, and the spatial pattern of size change is more complicated. Unlike the summer season, the duration and number of rainstorms also show strong changes in many regions. 
\label{figure:SpatialChange}
\end{figure}

We compare the baseline and future WRF runs using the same approaches described above.
The precipitation responses to climate change again differ by season, and show similar distinctions between summer convective and winter large-scale precipitation as those seen in the model-observation comparison.
In the summer, the main drivers of changing precipitation patterns are again rainstorm intensity and size (Table \ref{table:SummerChange} and Figure \ref{figure:SpatialChange}, first column). Changes in the other aspects (duration and the number of storms) are relatively small and not statistically significant. For convective precipitation, at least in the WRF model runs used here, the compensating mechanism that allows a 3\%/K increase in total precipitation amount that differs from the 6\%/K increase in precipitation intensity is a decrease in the size of individual rainstorms (-3\%/K). 

In the winter, changes in precipitation patterns in future climate conditions are more complicated (Table 4 and Figure \ref{figure:SpatialChange}, second column). Across the whole U.S.\, the winter intensity effect is about a third less than that in summer, but still significant (4.4\%/K) and larger than the change in total rainfall (2.5\%/K). In contrast to summer, reduction in the size, duration and number of storms all play a role in compensating for that discrepancy. (Size is estimated as the most important factor, but these differences are not statistically significant.) Interestingly, while each of these factors show different geographic patterns of change, the change in seasonal mean is much more spatially uniform.  
This regional variation in compensation mechanisms means that different regions show different spatio-temporal precipitation changes that can  be significant for socioeconomic impacts. 

Winter patterns of total precipitation changes are also less uniform than those in summer. In both seasons, a part of the U.S.\ Southwest shows decreased rather than increased precipitation under future climate conditions, but the area of drying becomes larger in the winter, extending to about 1/3 of the contiguous U.S.\ This area provides a useful test area for examining the mechansisms driving changes in rainstorm characteristics, since it presents an even stronger discrepancy between changes in intensity and total amount of precipitation than does the average U.S.\ In the drying region, the intensity response is muted (and even negative in places), but still on average positive (Tables \ref{table:SummerChangeDry} and \ref{table:WinterChangeDry}, the first and second rows). That is, average storm intensity increases (by +2-3\%/K in summer and winter) while total precipitation actually decreases (by -5-6\%/K). We therefore repeat the analysis of changes in precipitation characteristics on the region of drying. (The analyzed areas for each season roughly coincide with the red regions in the maps in the first row of Figure \ref{figure:SpatialChange}; see Figure S4 for the exact areas analyzed.) We find that reduction of rainstorm size remains the dominant compensating factor in both summer and winter, with storm shrinkage substantial enough (more than -6\%/K in both seasons) to result in decreased total precipitation even though storms are stronger (Tables \ref{table:SummerChangeDry} and \ref{table:WinterChangeDry}). The main drivers of changing precipitation patterns in drying areas again appear to be partially compensating changes in intensity and size.

Finally,  we conduct a preliminary graphical exploration of potential changes in rainstorm trajectories in changing climate conditions. Once rainstorms are identified across time, the trajectory of each storm is easily found by linking its central locations at each time step. Figure S5 in the Supplemental Material shows rainstorm trajectories in the baseline and the future model runs. Although more detailed analysis would be useful, we see no clear sign of changes. %(Need to say more) 
That is, regional differences in total precipitation change do not appear to be driven by deviations on storm trajectories. 

\section{Simulating Future Precipitation Patterns by Changing Size and Intensity Distributions} \label{section:simulation}

The WRF runs used analyzed here suggest that rainstorm properties would change siginficantly under future climate conditions, but %the WRF-Stage IV comparison also suggests that 
model biases are strong enough that model projections alone are unsuitable for impacts assessments. %would lead to misleading results. Many other studies have also cautioned against directly using model projections of future precipitation in climate risk assessments \citep[e.g.][]{ines2006bias,baigorria2007assessing,teutschbein2012bias,muerth2013need}.  To allow our findings to inform climate risk assessment studies, we need 
We therefore seek to produce simulations of future precipitation that reflect model-projected changes in rainstorm characteristics, but also incorporate information from observational data.

For simulating future rainstorms, attempting to correct underlying biases in model projections is not an appropriate strategy, because model biases in rainstorm characteristics are much larger than the expected changes from present to future climate states.
A better approach is that of ``data-driven simulations'', i.e.\ transforming present-day observations into a future simulation using information from climate model runs. The approach has been used in simulations of temperature \citep[e.g.][]{ho2012calibration,hawkins2013calibration,leeds2015simulation,poppick2015temperatures} and in a few studies of precipitation \citep[e.g.][]{hay2000comparison,raisanen2013projections,raty2014evaluation}, but in no case allowing for changes in spatio-temporal dependence. (This weakness is shared with bias correction approaches applied to individual time series.)  
%As we have seen in Section \ref{section:results}, the differences in rainstorm characteristics between the baseline WRF model run and Stage IV data are much larger than those between the baseline model run and future model run.
 % As a result, methods based on correcting model output need to make much more drastic and complicated changes  than data-driven simulation approaches. 

In this section we propose an algorithm for precipitation simulations that modifies the two aspects of rainstorm characteristics that change most significantly in future conditions, size and intensity. 
%We transform the observed precipitation record into a future precipitation simulation by changing the rainstorm size and intensity distributions as suggested by our model runs. 
This method should be highly appropriate for simulating future summer precipitation, when factors other than size and intensity are largely unchanged. For winter precipitation, it may be problematic because duration and number of storms also show large, generally compensating changes.
(Accounting for those changes would be complicated, because the number of storms increases in future projections in certain regions, requiring the creation of rain events not present in observations.)
Our simulation approach consists of two steps:
\begin{itemize}
\item[A.] Changing the size of individual rainstorms in the observational record  %(Subsection \ref{subsection:ChangingSize})
\item[B.] Changing the distribution of precipitation intensity in the transformed data from step 1 %(Subsection \ref{subsectionChangingEachCell})
\end{itemize}
{\color{black} In both cases, we account for regional differences in projected rainstorm changes using location-dependent resizing factors and grid-cell-specific intensity distribution transformations. (See Subsections 6a and 6b below for further details.)}
In the remainder of this section, we first describe the algorithm, and then validate it by testing whether it produces improved simulation results over a simple grid-cell level simulation that incorporates no spatial information. %in Subsection \ref{subsection:ComparingResults}. 
We describe only the basic steps here, and give details and equations in Section S6 in the Supplemental Material.

\subsection{Changing Individual Rainstorm Sizes} \label{subsection:ChangingSize}

\begin{figure}
\begin{center}
\includegraphics[scale=0.5]{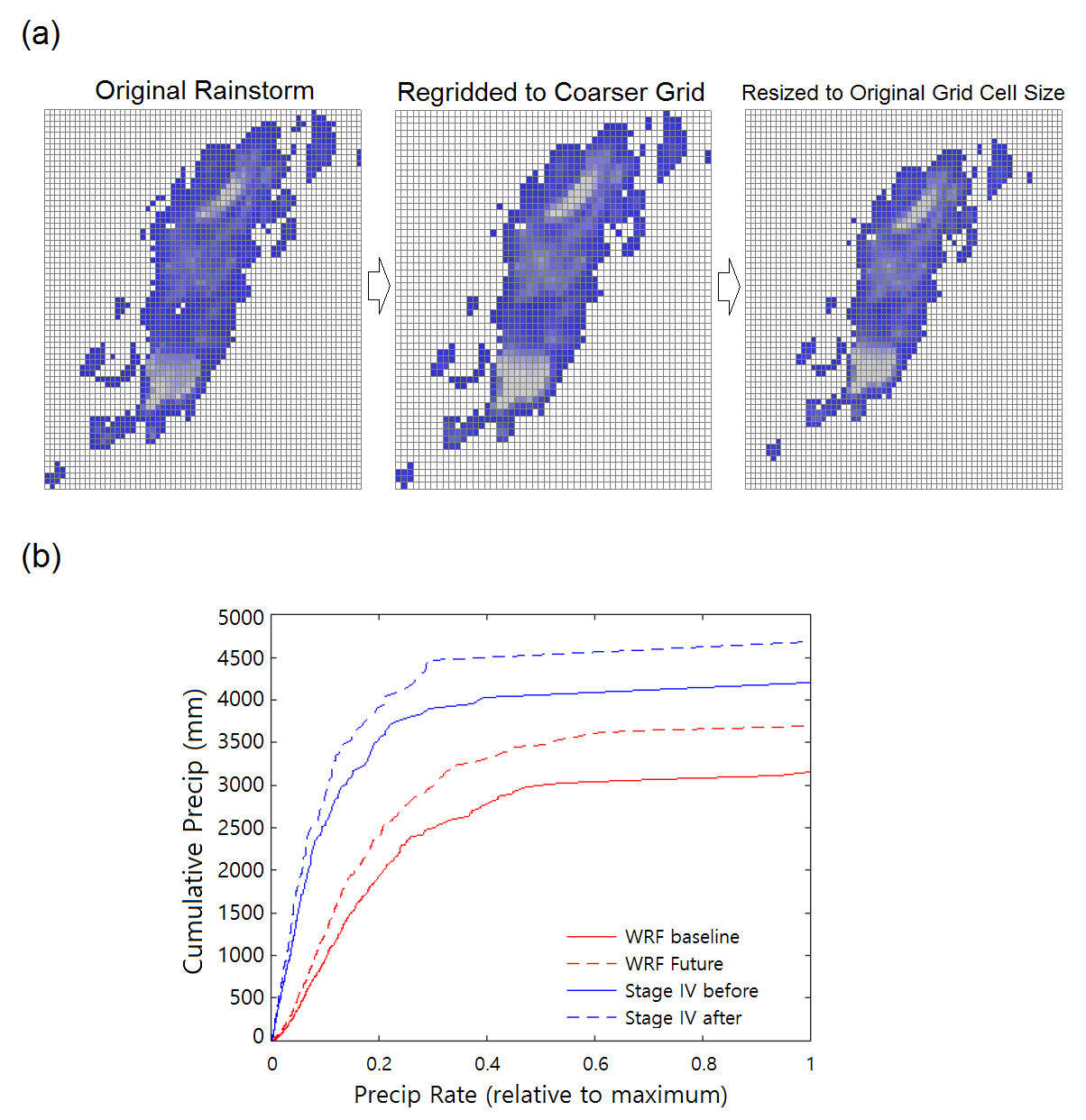}\\
\end{center}
\caption{Illustration of the two steps of our simulation algorithm. (a) Shrinking an observed rainstorm. We first remap each rainstorm event onto a coarser grid with spacing determined by a locally-determined size factor, and then resize to the original grid spacing. This approach resizes a rainstorm without changing its shape. (b) Transforming the overall intensity distribution at a single grid cell.
We rescale each quantile using a factor determined by the model projected changes (red).
 The plot shows the effect of this transformation applied to Stage IV data (blue).}
\label{figure:simulation}
\end{figure}

To change the sizes of observed rainstorms, we first determine the location-dependent resizing factors by comparing the estimated size functions between the baseline and future model runs (computed using the approach in Section 4\ref{section:SpatialAnalysis}).
We then resize each observed rainstorm at each time step using the resizing factor that corresponds to its central location. To avoid changing the shape of a storm, we follow a simple resizing procedure that involves changing the distances between the storm center and all individual grid cells in the storm by the same factor. %This procedure conceptually provides a simple way of changing rainstorm sizes, but applying it to observed rainstorms requires additional considerations because the resulting resized rainstorm needs to be on the same grid as the original rainstorm. 
We regrid the rainstorm mathematically to a grid whose spacing is the inverse of the resizing factor (Figure \ref{figure:simulation}a, second panel). % XX Won check that that's right. 
We then resize these new grid cells to those of the original grid, while keeping the center of the resulting rainstorm as close as possible to its original location (Figure \ref{figure:simulation}a, third panel). 

Note that in some cases a rainstorm event is split into two or more sub-storms that are far from each other. For resizing purposes, we consider that those two parts warrant individual treatment if the shortest distance between their edges is greater than 120 km. In this case we define separate central locations and separate resizing factors for each sub-storm in the scattered rainstorm event.

\subsection{Changing Intensity Distribution at Each Grid Cell Location} \label{subsectionChangingEachCell}

%The second step of our algorithm is to change the individual time series at each grid cell location generated by the resized rainstorms. The basic idea is to change each observed time series using the transformation given by the baseline and future model time series at the closest model grid cell. Since we are dealing with the time series given by the transformed observed rainstorms, we also use the baseline time series given by the baseline rainstorms which are transformed in the same way as the observed rainstorms. 
In the second step of the algorithm, we change the time series at each individual grid cell to adjust the marginal distribution of intensities. (See Figure \ref{figure:simulation}b for illustration.) The basic idea is to change the observed time series using a transformation that turns the marginal distribution of the model baseline time series into that of the corresponding model future time series. Since the observational data are not on the same grid as the model output, we determine each transformation based on the closest model grid cell location. %Note that we are transforming the observed time series from the dataset given by the resizing procedure described in the previous section, not the original observational dataset. 
%Therefore, to find the right transformation, we need to transform the baseline model output with the same resizing procedure and use the time series from the changed output.
To ensure that the model baseline output is treated identically to the observational data when adjusting intensities,
 we begin by applying the same resizing procedure to the baseline model output as used on the data. 

We then change the number of wet time steps (time steps with positive precipitation amount) at each location in the observational time series as suggested by the model at the same geographical location. 
If the model projection shows a decrease in wet time steps, we apply the same fractional change by turning the lowest observational intensities into zeros. If the number of wet time steps increases, we promote some dry time steps (time steps with no precipitation amount) to wet time steps to apply the same fractional change. To choose the dry time steps to promote, we use an idea similar to the protocol described in \citet{vrac2007statistical}, which creates rainfalls close in space or time to existing rainfalls. We first promote as many time steps as possible based on the precipitation amounts in their spatially adjacent grid cells. If there are not enough grid cells with positive precipitation in spatially adjacent grid cells, we select the time steps to promote based on the precipitation amounts in their temporally adjacent time steps. 

Once the number of wet time steps is changed, we transform the marginal distribution of positive intensities for each time series. For each value of an individual time series, we take as the rescaling factor the ratio between the corresponding quantiles of the baseline and future intensities. Multiplying each value in the time series by the appropriate rescaling factor produces a simulated time series for a single location. The process is then repeated across all grid cells to produce the simulation of future precipitation. 

\subsection{Comparing Simulation Results to Grid Cell-wise Simulation} \label{subsection:ComparingResults}

We evaluate the performance of our simulation approach by applying it to the baseline model run itself, whose future state is known. We then check if the resulting simulated precipitation patterns indeed reproduce statistical characteristics of the actual future run.
{\color{black} 
%XX We caution that test results presented here should not be over-interpreted as an evidence that our simulation based on the particular WRF model runs can accurately predict the future precipitation changes. 
This test can provide at least minimal validation that the algorithm accurately reproduces the changes in the WRF model runs it was trained on, and that it does so better than simpler alternative approaches. If the algorithm were to be used as an emulator, it would be appropriate to conduct a more challenging test, applying the algorithm to different runs of the same model not included in the original training set. (Our model simulations however preclude this as they consist of a single baseline and future realization.}  %Reviewer1 General 1
 
As a metric for comparison, we evaluate the distribution of ``dry spell'' lengths (successive timesteps without rainfall), a characteristic that is not itself explicitly adjusted in our simulation method. We compare the performance of our simulation, which incorporates spatial information about rainstorm events, to a simpler ``grid cell-wise simulation'' that does not use information from adjacent grid cells, but instead simply adjusts the intensity distribution at each location as in Section 6\ref{subsectionChangingEachCell}. Table \ref{table:KLDivergence} summarizes the results for five regions of the United States  (Midwest, Great Plains, Northeast, Wet South, and Dry South, see Figure S6 for region definitions). In most regions, the rainstorm event-based approach offers advantages over the simpler grid-cell-based approach in summer (when precipitation changes are dominated by intensity and size, the two factors our approach addresses). In winter, when rainstorm changes include features not captured by our approach (changes in number and duration), the event-based approach offers comparable performance but no clear advantage. 

%For the summer season, the results show that our approach produces a distribution of dry spell lengths that are closer to those from the future model run than the baseline model run. Moreover, our approach outperforms the grid cell-wise approach in terms of reproducing the distribution of dry spell lengths in the actual future run in most regions. 

\begin{figure}
\begin{center}
\includegraphics[scale=0.65]{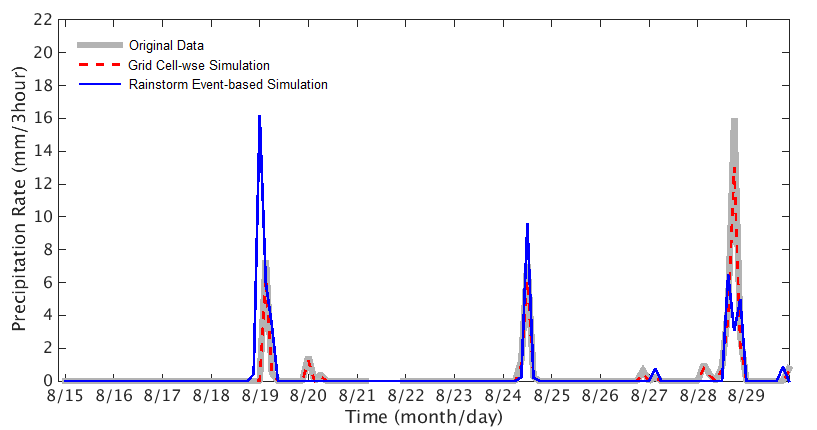}
\end{center}
\caption{
Illustration of differences in simulated future time series produced by our event-based algorithm (blue) and by the simpler grid-cell-wise approach described in Section \ref{section:simulation}c that uses no spatial information (dashed red). Both algorithms are applied to Stage IV observations (grey). The intent is to capture model-projected changes. 
The grid-cell-wise approach only rescales the intensity of the original time series. The event-based approach produces more diverse changes: it removes some rainstorm events completely (8/20 and 8/28) and drastically alters the local results of others (8/19 and 8/29).}
\label{figure:SimulExample}
\end{figure}

%The findings here show that changing the sizes of rainstorm events in our simulation approach leads to a better performance in simulating individual time series for the summer season. 
It may seem counterintuitive that incorporating spatial information is important for capturing dry spell lengths: 
at first glance, the event-based approach may seem relevant mostly for capturing the spatial characteristics of precipitation and not for improving the temporal distribution at an individual location. 
However, changing rainstorm events can lead to better simulation results even at individual grid cells by allowing more flexible changes. Figure \ref{figure:SimulExample} illustrates this point using an example time series of observational data at a location close to Chicago. Our rainstorm event-based approach, which alters sizes of events, often removes a whole precipitation event from the simulated time series if that event represents the edge of a larger system that is made smaller in the future simulation. %and therefore leads to a more flexible adjustment to the overall precipitation pattern at the location.  
The grid-cell level approach, by contrast, can adjust the magnitude of the precipitation rate and can reduce the duration of events, but cannot account for the changes in event occurrence that result when storms shrink. 

%Our response: We agree that we need to caution the readers about the limitations of this in-sample test (i.e.\ verification using the same dataset used to train the algorithm), and have added a few sentences to that effect (lines XX-XX). The test is a minimal test of the utility of the transformation algorithm: at absolute minimum, the algorithm should be able to reproduce the changes in the actual model runs it was trained on. It would be more challenging, and more valuable, to apply an algorithm developed from one model run to different runs of the same model. That is the standard means of evaluating a model emulation algorithm, and had we had more simulations we would have followed that practice here.  We have now explicitly discussed this in the text.  
 
\section{Discussion} \label{section:discussion}

% XX Summary paragraph of what you found
% XX include a bit on model-obs and on present-future
Climate model projections robustly imply that the spatio-temporal characteristics of precipitation events must change in future climate states. To help understand those potential changes, we have developed a new framework for analyzing changes or differences  
in rainstorm characteristics, including metrics for average intensity, size, duration, and number.
 The analysis framework is applicable both for comparison of future to present-day model simulations and for characterizing and validating the performance of models against observations. The same metrics allow us also to construct a method for simulating future precipitation events that combines model-projected changes with observational data.
 
Using this framework, we compare rainstorm properties in present and future high-resolution (12 km) dynamically downscaled model runs (WRF driven by CCSM4), and between those runs and the Stage IV radar-based observational data product. 
In all cases, the largest factors driving differences in rainstorm properties are intensity and size. In the summer season, when U.S.\ precipitation is predominantly convective, intensity and size are virtually the only factors of importance. In the model-observation comparison, WRF summer storms are too large but also too weak (leaving total precipitation consistent with observations). In the present-future model run comparison, WRF summer storms become smaller and stronger under future climate conditions (allowing total precipitation to rise less steeply than storm intensity). %That is, the main compensating mechanism for the discrepant rates of rise of intensity and total precipitation is a reduction in rainstorm size. 
The same size-intensity tradeoffs are apparent in winter, when U.S.\ precipitation tends to be large-scale, but differences in the duration and frequency of rainstorms also become important.  
In the model-observation comparison, WRF winter storms are not only too large and weak, but also too few and too long-lasting. In the model present-future comparison, WRF winter storms become smaller and stronger, as in summer, but also less numerous and of shorter duration. 
  
%It is potentially informative that in both summer and winter seasons, the projected future changes in rainstorm characteristics lie in all cases in the direction that would make the model more consistent with present-day observations. 
 These parallels may aid in understanding the underlying causes of model bias. In the WRF simulations analyzed here, model-observation biases are generally larger than the projected changes under nearly a century of business-as-usual CO2 emissions. (To compare model projections to model bias, see Tables S1 and S2, which reproduce Tables \ref{table:SummerDisc} and \ref{table:WinterDisc}, but for the Stage IV data area only.) The WRF future simulation projects that rainstorms become 20-26\% more intense (in summer and winter, respectively), but those future model storms remain weaker than observed present-day storms (by 13 and 29\%). Size biases are even more persistent: for example, future WRF winter rainstorms become 4\% smaller, but remain 130\% larger than in observations (Tables S1 and S2).
% XX need to revise numbers to ensure they are comparable, point reader to supp docs tables if the comparable numbers go only there. Dito for the Results section. or at least alert them to differences in area used.
The scale of these spatio-temporal biases, in one of the most commonly-used models for regional simulations, suggests that identifying detailed rainstorm characteristics is essential for validating and improving the representation of precipitation in models. 
%The method we present is designed for that purpose, since it is broadly applicable to all precipitation events, unlike current storm-identification approaches that focus highly-intense storm events in the context of weather forecasting \citep{jankov2005impact,davis2006object1,clark2009comparison,skamarock2008time}. 

%Model bias in rainstorm characteristics does not invalidate the utility of the present-future model comparison. 
Model bias in rainstorm characteristics, while a cause for concern, does not necessarily invalidate the utility of the present-future model comparison. The divergent changes in precipitation intensity and amount are well-understood consequences of physical constraints and are robust across models \citep{knutson1995time,allen2002constraints,held2006robust,willett2007attribution,stephens2008controls,wang2012review}. 
The rise in intensity is driven by the increased water content in warmer air; the rise in total precipitation is the consequence of a more infrared-opaque atmosphere, which mandates a greater export of energy from surface to atmosphere as latent heat. % XX should put citations here? 
%The resulting changes in intensity and amount are robust across models  and are replicated in the dynamically downscaled runs here. 
The robustness of these effects in models suggests that the same changes will manifest in the real world as climate evolves. Precipitation events in the real world would then also experience some compensating change in spatio-temporal characteristics. Model simulations provide a self-consistent system governed by these fundamental constraints and at least a potential analogue for understanding the real-world response. 

In the WRF runs analyzed here, compensation largely occurs through reduction in rainstorm size, with the size factor overwhelmingly dominant in summertime convective precipitation.  %The size-intensity tradeoff may reflect a fundamental mode of response of convection; these results point the way to future research directions. 
%The changes in summer and winter precipitation characteristics would have different implications for the societal impacts of changing hydrology. The robust rise in precipitation intensity in model projections has prompted concern about increase in severe flooding events. The WRF model runs here suggest that risk may be lessened by a reduction in storm size. 
These changes would have implications for the societal impacts of changing hydrology. The robust increases in precipitation intensity in model projections has prompted concern about increased severe flooding events. The WRF model runs here suggest that risk may be lessened by reductions in storm size. 
%From the viewpoint of regional hydrology, more severe but smaller storms may cause less flooding than more severe but less frequent storms. 
At the scale of a drainage basin, summer precipitation per rain event in our simulations rises not as Clausius-Clapeyron but as the smaller rise in total rainfall amount.  
In winter, precipitation events show more complex and potentially more impactful changes than in summer. Intensity increases are somewhat less than in summer, but storm frequency (number of initializations) plays a larger role in compensating for the intensity/amount discrepancy. The combination means that at the scale of a drainage basin, precipitation per event would increase more in winter than in summer. In addition, wintertime precipitation changes in WRF are more spatially heterogeneous, implying that local impacts of future precipitation changes may be geographically diverse (which further limits the possbility of diagnosing aggregate effects through observational studies in limited regions; see e.g. \citet{berg2013strong}). % studied individual rain events in Germany in radar-based data and concluded that while convective precipitation increases in intensity at warmer temperatures, large-scale precipitation does not. The WRF model runs analyzed here conflict with that finding, but also suggest that a possible explanation may simply be geographic diversity in response. 
Given the societal importance of projected changes in spatio-temporal precipitation characteristics, a research priority should be understanding whether the response shown in these WRF simulations is robust across models.
All moodel simulations must involve some change in precipitation characteristics that compensates for diverging changes in intensity and amount, but different models may ``find'' different solutions to those constraints. Analysis methods that identify individual precipitation events are essential for this purpose. 

% because hydrology matters, important to simulate accurately for impacts assessments.
The societal importance of potential flood increases also motivates our development of precipitation simulation methods. Future projections must play a role in risk assessments, but the large observation-model biases we have demonstrated mean that even if model-projected changes are believed to be true, risk assessments should not use model output alone. Given the scale of the biases, simple bias correction methods are also not appropriate. We therefore advocate instead a data-driven approach that extracts only projected \emph{changes} from model runs and use them to transform observations into future projections. We have shown that our event-based simulation of WRF outperforms simpler simulation methods for summer precipitation, where compensation occurs largely through reduction in rainstorm size. Simulating WRF winter precipitation (or precipitation in any model simulation with more complex changes), remains an outstanding research question. Since our WRF runs project a future increase in winter storm number {\color{black}in certain regions (e.g.\ parts of the East Coast and Midwest, and most of the Southeast)}, a data-driven simulation algorithm could not capture such an increase by simply modifying existing storms, and would have to  
create new rainstorms in the observational record. Creating physically plausible and {\color{black} spatially consistent} new precipitation events would likely require consideration of many other variables, including temperature and relative humidity. %Even in the summer, the simulation methods we have described may not be sufficient for all purposes: agricultural assessments would require joint simulation of precipitation and temperature. 

% Other future directions
This analysis represents, to our knowledge, a first attempt to understand and simulate model-projected changes in precipitation characteristics such as size, duration, and frequency of rainstorm events. The methodology for identifying and tracking storms should open up many other potential areas of research.  While we have analyzed only a single model, multi-model comparison is a clear research priority. 
It would also be useful to subdivide precipitating events 
 according to meteorological context, and to separately examine the behavior of individual convective cells. 
Another potentially important area is understanding future changes in storm tracks. While our preliminary analysis shows no obvious changes in rainstorm trajectories, further analysis would be useful.
The existing studies that examine rainstorm trajectories do so rather informally \citep{hodges1994general,morel2002climatology2,xu2005kernel,cressie2012dynamical} and formal statistical analysis of rainstorm trajectories is a largely unexplored area. 
Finally, studies at other latitudes may also provide additional insight. The changes in midlatitudes precipitation characteristics shown here are not a simple function of temperature rise and would not be identical across geographic regions. In some parts of the tropics, total precipitation may rise \textit{more} rather than less steeply than Clausius-Clapeyron, requiring a different compensating response. Understanding how precipitation characteristics change in response to geographically differing constraints on the hydrological cycle may provide new insight into convective organization and structure. All these studies are made possible only given a methodology for identifying and studying the physical characteristics of individual storms.
\section*{Acknowledgments}
The authors thank Dongsoo Kim and Mihai Antinescu for helpful comments and suggestions. This work was conducted as part of the Research Network for Statistical Methods for Atmospheric and Oceanic Sciences (STATMOS), supported by NSF awards \#1106862, 1106974, and 1107046, and the Center for Robust Decision-making on Climate and Energy Policy (RDCEP), supported by the NSF Decision Making Under Uncertainty program award \#0951576.

%%%%%%%%%%%%%%%%%%%%%%%%%%%%%%%%%%%%%%%%%%%%%%%%%%%%%%%%%%%%%%%%%%%%%
% APPENDIXES
%%%%%%%%%%%%%%%%%%%%%%%%%%%%%%%%%%%%%%%%%%%%%%%%%%%%%%%%%%%%%%%%%%%%%

%% If more than one appendix, use \appendix[<letter>], e.g.,
% \appendix[A] 

%\appendixtitle{Title of Appendix}

%\subsection{Appendix section}

%%%%%%%%%%%%%%%%%%%%%%%%%%%%%%%%%%%%%%%%%%%%%%%%%%%%%%%%%%%%%%%%%%%%%
% REFERENCES
%%%%%%%%%%%%%%%%%%%%%%%%%%%%%%%%%%%%%%%%%%%%%%%%%%%%%%%%%%%%%%%%%%%%%
% This shows how to enter the commands for making a bibliography using
% BibTeX. It uses references.bib and the ametsoc2014.bst file for the style.

 \bibliographystyle{ametsoc2014}
 \bibliography{short,references}

\clearpage
%%%%%%%%%%%%%%%%%%%%%%%%%%%%%%%%%%%%%%%%%%%%%%%%%%%%%%%%%%%%%%%%%%%%%
% TABLES
%%%%%%%%%%%%%%%%%%%%%%%%%%%%%%%%%%%%%%%%%%%%%%%%%%%%%%%%%%%%%%%%%%%%%

\begin{table}
\begin{center}
Model vs. Observations, Summer (JJA)\\
\begin{tabular}{l|r|r|r}
\hline
&   Observation & Model &  Difference (\%)  \\
 \hline
Seasonal Mean (cm/season) & 21 & 20& -4.7 \\
\hdashline
Intensity (mm/hour) &  3.8 & 2.6 &  -33   \\ 
Size (10$^4$ km$^2$) &  3.4 & 5.3 & 51   \\ 
Duration (hour) & 10.9 & 9.7 & -11  \\
Number of Storms (storms/hour)& 1.5 & 1.6 & 5.1 \\
\hline
\end{tabular}
\end{center}
\caption{Comparison of summer rainstorm event characteristics between the WRF model output and Stage IV observational data.
First and second columns show mean values for model and observations and the third column their fractional differences. For consistency we include only data east of $114^\circ$ W, excluding the region where Stage IV data are problmatic. 
The analysis fully decomposes differences in total precipitation into differences in the four factors defined in Section 4\ref{section:factorization} (intensity, size, duration, and number).
While overall precipitation in model and observations are similar (top row), model output shows large counteracting biases, especially in intensity and size: model rainstorm events are too large and too weak. 
The number of rainstorm initializations in the last row is the variable $N$ in Section \ref{section:RainstormAnalysis1}.  % could delete this last sentence if it's not needed.
}
\label{table:SummerDisc}
\end{table}
\clearpage
\begin{table}
\begin{center}
Model vs. Observations, Winter (DJF)\\
\begin{tabular}{l|r|r|r}
\hline
&  Observation & Model &  Difference (\%) \\
 \hline
Seasonal Mean (cm/season)& 13& 14 & 14\\
\hdashline
Intensity (mm/hour) &  2.4 & 1.4 &  -39  \\ 
Size (10$^4$ km$^2$) & 9.9 & 25& 150  \\ 
Duration (hour) & 15 & 24 & 61  \\
Number of Storms (storms/hour) & 0.35 & 0.17 & -53\\
\hline
\end{tabular}
\end{center}
\caption{As in Table 1 but for the winter season. As in summer, large counteracting biases result in roughly similar  total precipitation in model and observations. Biases are more complex in winter than summer: model winter precipitation events are weaker, larger, longer, and less frequent than in obserations. Size remains the most discrepant characteristic, with winter bias substantially larger than in summer (150\% vs.\ 51\%).}
\label{table:WinterDisc}
\end{table}
\clearpage

\begin{table}
\begin{center}
Baseline vs. Future, Summer (JJA), Contiguous U.S.
\begin{tabular}{l|r|r|r|r}
\hline
&  Baseline & ~~Future~~ &  Change (\%/K) & 95\% CI for Change \\
 \hline
Seasonal Mean (cm/season)&  21.0 &   24 &    3.0 &    (0.98, 5.1)\\
\hdashline
Intensity (mm/hour) &  2.4 & 3.1 &  6.2  & (3.9, 8.4) \\ 
Size (10$^4$ km$^2$) &  4.9& 4.3 & -3.2  & (-5.2, -0.70) \\ 
Duration (hour) & 9.8 & 9.9 & 0.17  & (-0.26, 0.52)) \\
Number of Storms (storms/hour) & 1.9 & 2.3 & 0.53  & (-0.35, 1.4)\\
\hhline{=====}
Temperature (K) & 295.8 & 300.1 & 4.3 & \\
\hline
\end{tabular}
\end{center}
\caption{Changes in summer rainstorm characteristics between model baseline and future runs. Precipitation changes are decomposed into factors as in Tables 1 and 2, but here we use all data over the study area. First and second columns show mean values in baseline and future runs, third column the fractional changes, and fourth column the 95\% confidence intervals (see Section 4\ref{section:ComparisonMethod}) for those changes. Changes are normalized by the regional mean temperature increase (4.3K), shown in last row. 
Increases in seasonal mean precipitation (3.0\%/K, or $\sim$12\% in total) and intensity (6.2\%/K, or $\sim$27\% in total) are consistent with existing studies. Both these changes are statistically significant (i.e.\ lower limits are greater than 0\%/K). Compensation for these divergent changes is provided primarily by the size factor, which significantly decreases (-3.2\%/K, or -13.7\% in total). Changes in duration and frequency of storms are not statistically significant.}
 \label{table:SummerChange}
\end{table}

\clearpage
\begin{table}
\begin{center}
Baseline vs. Future, Winter (DJF), Contiguous U.S.
\begin{tabular}{l|r|r|r|r}
\hline
&  Baseline & ~~Future~~ &  Change (\%/K) & 95\% CI for Change \\
 \hline
Seasonal Mean (cm/season)& 20& 22 & 2.5 & (-1.2, 6.6) \\
\hdashline
Intensity (mm/hour) &  1.6 & 1.9 &  4.4  & (2.5, 5.8) \\ 
Size (10$^4$ km$^2$) &  30 & 29 & -0.76  & (-3.1, 2.0) \\ 
Duration (hour) & 25 & 25 & -0.34  & (-1.6, 1.5) \\
Number of Storms (storms/hour) & 0.17 & 0.17 & -0.73  & (-2.5, 1.0)\\
\hhline{=====}
Temperature (K) & 274.9  & 279.5  & 4.6 \\
\hline
\end{tabular}
\end{center}
\caption{As in Table 3 but for the winter season. Increases in mean precipitation and intensity are somewhat lower than summer but again discrepant at 2.5\%/K (11\% in total) and 4.4\%/K (20\% in total), respectively. (Winter temperature rises slightly more than in summer, at 4.6 K.) 
Changes in other rainstorm characteristics are more complex than those in summer. The size decrease is smaller, with reductions in storm number and duration of similar magnitude. Uncertainties are sufficiently large that of all factors, including total precipation, only the intensity change is statistically significant.  % XX big question on what CIs mean for seasonal mean, whether this is the right way to talk about them
}
 \label{table:WinterChange}
\end{table}
\clearpage

\begin{table}
\begin{center}
Baseline vs. Future, Summer (JJA), Dry Region
\begin{tabular}{l|r|r|r|r}
\hline
&  Baseline & ~~Future~~ &  Change (\%/K) & 95\% CI for Change \\
 \hline
Seasonal Mean (cm/season)& 3.2 & 2.6 & -4.6 & (-9.5, 2.0) \\
\hdashline
Intensity (mm/hour) &  2.8 & 3.1 &  2.4  & (-1.0, 5.0) \\ 
Size (10$^4$ km$^2$) &  3.8& 2.7 & -6.6  & (-9.9, -2.5) \\ 
Duration (hour) & 8.6 & 8.3 & -0.8  & (-1.7, 0.0)) \\
Number of Storms (storms/hour) & 0.39 & 0.41 & 1.2  & (-0.70, 3.4)\\
\hhline{=====}
Temperature (K)  & 299.5  & 304.0  & 4.5  \\
\hline
\end{tabular}
\caption{As in Table 3 but here considering only the regions where summer mean precipitation decreases. (See Figure S4 in the Supplemental Material.) Summer temperature rise is similar to that over the entire study area (4.5K vs.\ 4.3K). The reduction in mean precipitation is largely explained by a lesser increase in intensity than in the overall study area and a larger decrease in the size factor.}
 \label{table:SummerChangeDry}
\end{center}
\end{table}
\clearpage

\begin{table}

\begin{center}
Baseline vs. Future, Winter (DJF), Dry Region
\begin{tabular}{l|r|r|r|r}
\hline
&  Baseline & ~~Future~~ &  Change (\%/K) & 95\% CI for Change \\
 \hline
Seasonal Mean (cm/season)& 4.6 & 3.4 & -6.0 & (-10, -0.63) \\
\hdashline
Intensity (mm/hour) &  1.7 & 1.9 &  3.1  & (-0.26, 5.3) \\ 
Size (10$^4$ km$^2$) &  19 & 14 & -5.9  & (-9.3, -3.2) \\ 
Duration (hour) & 22.0 & 20.2 & -1.4  & (-3.0, 0.31)) \\
Number of Storms (storms/hour) & 0.067 & 0.066 & -0.50 & (-3.3, 2.5)\\
\hhline{=====}
Temperature (K)  & 278.2 K & 282.6 K & 4.4 K \\
\hline
\end{tabular}
\end{center}
\caption{As in Table 5 but here for the winter season. The region where precipitation decreases is larger in winter than in summer; see Figure S4 in the Supplemental Material. Winter temperature rise is similar to that over the entire study area (4.4K vs.\ 4.6K), but the region shows a lesser increase in rainstorm intensity and, in contrast to the entire study area, a clear and statistically significant decrease in the size factor.}
 \label{table:WinterChangeDry}
\end{table}
\clearpage
\begin{table}
\begin{center}
KL-divergence vs. Actual Future (Dry spell Length)\\
~~~~~~~~~~~~~~~~~~~~~~~~~~~~~~~~~~~~~~~~~~~~~~~~~~~~~~~~~~~~~~~~~~~~~~~~~~~~~~~~~~~~~~~~~~~~($ \times10^{-2}$)\\
\begin{tabular}{l|l|ccccc}
\hline
Season&Region&i. Baseline&ii. Grid cell-wise&iii. Storm-based&i-iii&ii-iii\\
\hline
\multirow{ 5}{*}{Summer}& Wet South&   0.47 &   0.17 &   0.09 &   0.38 &   0.07\\
& Dry South &   0.49  &  0.25  &  0.22  &  0.26 &   0.03\\
& East Coast &   0.39 &   0.37  & 0.28  &  0.11 &   0.09\\
& Midwest &   0.44 &   0.18  &  0.12   & 0.32 &   0.07\\
& Great Plains &   0.19  &  0.11  &  0.11 &  0.08  &  0.00\\
\hdashline
\multirow{ 5}{*}{Winter}& Wet South&   0.41 &   0.33  &  0.33 &   0.09 &   0.00\\
& Dry South &   1.32 &   0.46 &   0.49 &   0.83&   -0.03\\
& East Coast &   0.29  &  0.28   & 0.28  &  0.01  &  0.00\\
& Midwest &   0.34  &  0.30 &   0.28 &   0.06 &   0.02\\
& Great Plains &   0.67 &   0.49  &  0.47 &  0.20 &   0.03\\
\hline
\end{tabular}
\end{center}
\caption{
Evaluating simulation algorithms by comparing distributions of dry spell lengths in the future model run to thoese in simulations of it that transform the baseline run. We compute Kullback-Leibler (KL) divergences between distributions in ten regional simulations and in the corresponding model output. 
In each case, we pool dry spell lengths from all grid cells to make a single histogram. %, then compare histograms for the simulation and future model run.  
 For reference, we first show (i) the KL divergence between the model future and unchanged baseline runs. As expected, divergences are large, reflecting the projected precipitation changes. %A simulation to provide value, it must produce lower KL divergences than these values. 
We then show divergences to simulations produced with (ii) the grid-cell-wise algorithm described in Section \ref{section:simulation}c and (iii) our proposed simulation algorithm.  Columns labeled i-iii and ii-iii show differences, to evalute relative performances. Postiive values in i-iii confirm that our event-baed simulation provides value in capturing model-projected future changes to dry spell lengths. Positive values in ii-iii for summer cases show that our simulation outperforms the simpler grid-cell-wise approach. Near-zero values in ii-iii for winter cases mean that our event-based simulation offers no clear advantage (unsurprisingly, as it cannot capture changes in rainstorm duration and number).}
\label{table:KLDivergence}
\end{table}
\clearpage
%\begin{center}
%\begin{tabular}{c|r|r}
%\hline
%& Summer (JJA) & Winter (DJF)  \\
% \hline
%Overall Amount & -4.71 \%/K & -\textit{6.25}\% \\
%Intensity &  2.35\% & 3.00\% \\ 
%Size &  -\textit{6.75}\% & -\textit{7.2}\%   \\ 
%Duration & -\textit{0.67}\% & 0.05\% \\
%Rate& 1.24\%  & 0.38\% \\
%\hline 
%Temperature &4.5 K&4.4 K  \\
%\hline
%\end{tabular}
%\end{center}

%\begin{center}
%\begin{tabular}{c|r|r}
%\hline
%& Summer (JJA) & Winter (DJF)  \\
% \hline
%Overall Amount & \textit{4.32} \%/K & \textit{4.91} \%/K\\
%Intensity &  \textit{6.90} \%/K & \textit{4.89} \%/K\\ 
%Size &  -\textit{2.50} \%/K &  -0.08 \%/K\\ 
%Duration & \textit{0.46} \%/K  & 0.46 \%/K\\
%Rate& 0.10 \%/K & -0.76 \%/K\\
%\hline 
%Temperature& 4.3 K&4.6 K \\
%hline 
%\end{tabular}
%\end{center}

\clearpage

%%%%%%%%%%%%%%%%%%%%%%%%%%%%%%%%%%%%%%%%%%%%%%%%%%%%%%%%%%%%%%%%%%%%%
% FIGURES
%%%%%%%%%%%%%%%%%%%%%%%%%%%%%%%%%%%%%%%%%%%%%%%%%%%%%%%%%%%%%%%%%%%%%

\clearpage
% XX let's say less about mechanisms here to not get in trouble!
\clearpage

% To remove the Western region where Stage IV data are problematic, we exclude the model output and data from the region west of $114^\circ$ W from the comparison. 

\clearpage

% XX .. could also state 4 numbers for amount excluded in each case.

\clearpage

\clearpage

\clearpage

\clearpage

\clearpage

\clearpage

\clearpage
% Illustration of the two steps of our simulation algorithm. a) Shrinking an observed rainstorm. Example taken is a summertime convective storm from the Midwest (left). Rainstorm is resampled to a coarser grid according to the scaling factor (center), and then resized to original grid cell size (right). b) Transforming the intensity distribution. Example taken is a Midwest grid cell near Chicago; WRF is biased dry in this location. Precipitation intensities are adjusted.. (describe, maybe use your last sentence here). Plot here shows outcome expressed as cumulative precipitation amount vs precipitation rate. (then tell reader what to look at that’s interesting - maybe even though precip amounts are different in model and data, algortihm applies same fractional change).
\clearpage

\end{document}